\input harvmac
\input amssym.tex

\def\grtsim{\gtrsim}

\newcount\figno
\figno=0
\def\fig#1#2#3{
\par\begingroup\parindent=0pt\leftskip=1cm\rightskip=1cm\parindent=0pt
\global\advance\figno by 1
\midinsert
\epsfxsize=#3
\centerline{\epsfbox{#2}}
\vskip 12pt
{\bf Fig. \the\figno:} #1\par
\endinsert\endgroup\par
}
\def\figlabel#1{\xdef#1{\the\figno}}
\def\encadremath#1{\vbox{\hrule\hbox{\vrule\kern8pt\vbox{\kern8pt
\hbox{$\displaystyle #1$}\kern8pt}
\kern8pt\vrule}\hrule}}
\def\underarrow#1{\vbox{\ialign{##\crcr$\hfil\displaystyle
 {#1}\hfil$\crcr\noalign{\kern1pt\nointerlineskip}$\longrightarrow$\crcr}}}
%
\overfullrule=0pt

%
\def\tilde{\widetilde}
\def\bar{\overline}

\def\R{{\bf R}}

\font\zfont = cmss10 

\def\bigone{\hbox{1\kern -.23em {\rm l}}}
\def\ZZ{\hbox{\zfont Z\kern-.4emZ}}
\def\half{{\litfont {1 \over 2}}}

\def\Re{{\rm Re ~}}


\def\ls{\ell_s}

\def\gs{g_s}
\def\lp10{\ell_P^{10}}
\def\lp11{\ell_P^{11}}
\def\R11{R_{11}}

\def\GeV{{\rm GeV}}
\def\MeV{{\rm MeV}}
%
\def\a{\alpha}
\def\b{\beta}
\def\g{\gamma}
\def\d{\delta}
\def\e{\epsilon}

\def\m{\mu}
\def\n{\nu}
\def\la{\lambda}

\def\s{\sigma}
\def\om{\omega}

\def\G{\Gamma}

%
%
\def\l{\ell}
\def\ls{\ell_s}
\def\l{\left}
\def\r{\right}

\def\det{{\rm det}\,}
\def\tr{{\rm tr}\,}
\def\Tr{{\rm Tr}\,}
\def\tilde{\widetilde}

\def\p{\partial}

%
%

\def\QZ{\Bbb{Z}}

\def\QZ{\Bbb{Z}}
\def\p{\partial}

\def\half{{1\over 2}}
\def\Tr{{{\rm Tr~ }}}
\def\tr{{\rm tr\, }}

\def\Re{{\rm Re\hskip0.1em}}

\def\CL{{\cal L}}

\def\CQ{{\cal Q}}

\lref\AngelantonjHI{
  C.~Angelantonj, I.~Antoniadis, E.~Dudas and A.~Sagnotti,
  ``Type-I strings on Magnetised Orbifolds and Brane Transmutation,''
  Phys.\ Lett.\ B {\bf 489}, 223 (2000)
  [arXiv:hep-th/0007090].
}

\lref\AntoniadisDT{ I.~Antoniadis and S.~Dimopoulos, ``Splitting
Supersymmetry in String Theory,'' arXiv:hep-th/0411032.
}

\lref\AntoniadisIG{ I.~Antoniadis, N.~Arkani-Hamed, S.~Dimopoulos
and G.~R.~Dvali, ``New Dimensions at a Millimeter to a Fermi and
Superstrings at a TeV,'' Phys.\ Lett.\ B {\bf 436}, 257 (1998)
[arXiv:hep-ph/9804398].
}

\lref\banksone{T. Banks and M. Dine, "Couplings and Scales in
Strongly Coupled Heterotic String Theory," Nucl. Phys. {\bf B479}
(1996) 173, arXiv:hep-th/9605136.}

\lref\bankstwo{T. Banks and M. Dine, "The Cosmology of String
Theoretic Axions," Nucl. Phys. {\bf B505} (1997) 445,
arXiv:hep-th/9608197.}

\lref\bankscosmo{T. Banks, M. Dine and M. Graesser, "Supersymmetry,
Axions and Cosmology," Phys. Rev. {\bf D68} (2003) 0705011,
arXiv:hep-ph/0210256.}

\lref\bankslarge{T. Banks, M. Dine, P. J. Fox and E. Gorbatov, "On
the Possibility of Large Axion Decay Constants," JCAP {\bf 0306}
(2003) 001, arXiv:hep-th/0303252.}

\lref\BanksSG{
  T.~Banks and M.~Dine,
  ``Coping with strongly coupled string theory,''
  Phys.\ Rev.\ D {\bf 50}, 7454 (1994)
  [arXiv:hep-th/9406132].
}

\lref\BarrHK{ S.~M.~Barr, ``Harmless Axions In Superstring
Theories,'' Phys.\ Lett.\ B {\bf 158}, 397 (1985).
}

\lref\BergshoeffDE{
  E.~A.~Bergshoeff and M.~de Roo,
  Nucl.\ Phys.\ B {\bf 328}, 439 (1989).
}

\lref\CandelasHD{
  P.~Candelas, M.~Lynker and R.~Schimmrigk,
  ``Calabi-Yau Manifolds In Weighted P(4),''
  Nucl.\ Phys.\ B {\bf 341}, 383 (1990).
}

\lref\CascalesPT{ J.~F.~G.~Cascales and A.~M.~Uranga, ``Chiral
$4d$ String Vacua with $D$-Branes and Moduli Stabilization,''
arXiv:hep-th/0311250.
}

\lref\ChoiWV{ K.~Choi, ``Quintessence, Flat Potential and
String/$M$-Theory Axion,'' arXiv:hep-ph/9912218.
}

\lref\ChoiXN{ K.~Choi, ``String or $ M$-Theory Axion as a
Quintessence,'' Phys.\ Rev.\ D {\bf 62}, 043509 (2000)
[arXiv:hep-ph/9902292].
}

\lref\ConlonTQ{
  J.~P.~Conlon,
  ``The QCD Axion and Moduli Stabilisation,''
  arXiv:hep-th/0602233.
}

\lref\BalasubramanianZX{
  V.~Balasubramanian, P.~Berglund, J.~P.~Conlon and F.~Quevedo,
  ``Systematics of Moduli Stabilisation in Calabi-Yau Flux Compactifications,''
  JHEP {\bf 0503}, 007 (2005)
  [arXiv:hep-th/0502058].
}

\lref\DimopoulosKN{
  S.~Dimopoulos and S.~Thomas,
  ``Dynamical Relaxation of the Supersymmetric CP Violating Phases,''
  Nucl.\ Phys.\ B {\bf 465}, 23 (1996)
  [arXiv:hep-ph/9510220].
}

\lref\WenJZ{
  X.~G.~Wen and E.~Witten,
  Phys.\ Lett.\ B {\bf 166}, 397 (1986).
}

\lref\DineBQ{ M.~Dine, N.~Seiberg, X.~G.~Wen and E.~Witten,
``Nonperturbative Effects On The String World Sheet. 2,'' Nucl.\
Phys.\ B {\bf 289}, 319 (1987).
}

\lref\DineZY{ M.~Dine, N.~Seiberg, X.~G.~Wen and E.~Witten,
``Nonperturbative Effects On The String World Sheet,'' Nucl.\ Phys.\
B {\bf 278}, 769 (1986).
}

\lref\DenefDM{
  F.~Denef, M.~R.~Douglas and B.~Florea,
  ``Building a better racetrack,''
  JHEP {\bf 0406}, 034 (2004)
  [arXiv:hep-th/0404257].
}

\lref\FoxKB{
  P.~Fox, A.~Pierce and S.~Thomas,
  ``Probing a QCD string axion with precision cosmological measurements,''
  arXiv:hep-th/0409059.
}

\lref\GreenMN{
  M.~B.~Green, J.~H.~Schwarz and E.~Witten,
  ``Superstring Theory. Vol. 2: Loop Amplitudes, Anomalies And Phenomenology,''
}

\lref\gukov{B. S. Acharya and S. Gukov, "M theory and Singularities
of Exceptional Holonomy Manifolds," arXiv:hep-th/0409191.}

\lref\kaplun{V. S. Kaplunovsky, "One Loop Threshold Effects in
String Univfication," arXiv: hep-th/9205070, arXiv:hep-th/9205068,
Nucl. Phys. {\bf B307} (1988) 145, Erratum-ibid, {\bf B382} (1992)
436.}

\lref\kaplund{L. J. Dixon, V. Kaplunovsky, J. Louis, "Moduli
Dependence of String Loop Corrections To Gauge Coupling Constants,"
Nucl. Phys. {\bf B355} (1991) 649.}

\lref\LythTX{
  D.~H.~Lyth and E.~D.~Stewart,
  ``Constraining the inflationary energy scale from axion cosmology,''
  Phys.\ Lett.\ B {\bf 283}, 189 (1992).
}

\lref\MarchesanoXZ{ F.~Marchesano and G.~Shiu,
JHEP {\bf 0411}, 041 (2004) [arXiv:hep-th/0409132].
}

\lref\MarchesanoYQ{ F.~Marchesano and G.~Shiu,
Phys.\ Rev.\ D {\bf 71}, 011701 (2005) [arXiv:hep-th/0408059].
}

\lref\polchinski{J. Polchinski, {\it String Theory,} Vol. 2
(Cambridge University Press, 1998).}

\lref\PolchinskiRR{
  J.~Polchinski,
  ``String theory. Vol. 2: Superstring theory and beyond,''
}

\lref\romalis{ M. V. Romalis, W. C. Griffith, J. P. Jacobs, and E.
N. Fortson, ``New Limit on the Permanent Electric Dipole Moment of
${}^{199}Hg$,'' Phys. Rev. Lett. {\bf 86,} 2505 (2001).}

\lref\SchwarzMG{ J.~H.~Schwarz and A.~Sen, ``Duality symmetries of
4-D heterotic strings,'' Phys.\ Lett.\ B {\bf 312}, 105 (1993)
[arXiv:hep-th/9305185].
}

\lref\SchwarzVS{ J.~H.~Schwarz and A.~Sen, ``Duality symmetric
actions,'' Nucl.\ Phys.\ B {\bf 411}, 35 (1994)
[arXiv:hep-th/9304154].
}

\lref\SeckelTJ{
  D.~Seckel and M.~S.~Turner,
  Phys.\ Rev.\ D {\bf 32}, 3178 (1985).
}

\lref\SenVD{ A.~Sen, ``F-theory and Orientifolds,'' Nucl.\ Phys.\ B
{\bf 475}, 562 (1996) [arXiv:hep-th/9605150].
}

\lref\shellard{E. P. S. Shellard and R. A. Battye, "Cosmic Axions,"
arXiv:astro-ph/9802216.}

\lref\SrednickiXD{
  M.~Srednicki,
  ``Axion Couplings To Matter. 1. CP Conserving Parts,''
  Nucl.\ Phys.\ B {\bf 260}, 689 (1985).
}

\lref\StromingerKS{
  A.~Strominger,
  Phys.\ Rev.\ Lett.\  {\bf 55}, 2547 (1985).
}

\lref\tamar{T. Friedmann and E. Witten, "Unification Scale, Proton
Decay, and Manifolds of $G_2$ Holonomy," arXiv:hep-th/0211269.}

\lref\TurnerUZ{
  M.~S.~Turner and F.~Wilczek,
  ``Inflationary Axion Cosmology,''
  Phys.\ Rev.\ Lett.\  {\bf 66}, 5 (1991).
}

\lref\WenJZ{ X.~G.~Wen and E.~Witten,
Phys.\ Lett.\ B {\bf 166}, 397 (1986).
}

\lref\WeinbergKR{
  S.~Weinberg,
  ``The Quantum Theory of Fields. Vol. 2: Modern Applications,''
}

\lref\WeinbergMT{
  S.~Weinberg,
  {\it The Quantum Theory of Fields},  Vol. I (Cambridge University Press, 1995).}

\lref\WeinbergCR{
  S.~Weinberg,
  ``The Quantum Theory of Fields.  Vol. 3: Supersymmetry,''
}

\lref\WilczekCR{
  F.~Wilczek,
  ``A model of anthropic reasoning, addressing the dark to ordinary matter
  coincidence,''
  arXiv:hep-ph/0408167.
}

\lref\wittenan{E. Witten, "Anomaly Cancellation On $G_2$-Manifolds,"
arXiv:hep-th/0108165.}

\lref\wittenatiyah{M. Atiyah and E. Witten, "$M$ Theory Dynamics on
a Manifold of $G(2)$ Holonomy," Adv. Theor. Math. Phys. {\bf 6}
(2003) 1, arXiv:hep-th/0107177.}

\lref\wittenc{E. Witten, "On Flux Quantization in $M$-theory and
The Effective Action," arXiv:hep-th/9609122, J. Geom. Phys {\bf
22} (1997) 1.}

\lref\wittenflux{E. Witten, "On Flux Quantization in $M$ Theory and
the Effective Action," J. Geom. Phys. {\bf 22} (1997) 1,
arXiv:hep-th/9609122.}

\lref\WittenMZ{
  E.~Witten,
  ``Strong Coupling Expansion Of Calabi-Yau Compactification,''
  Nucl.\ Phys.\ B {\bf 471}, 135 (1996)
  [arXiv:hep-th/9602070].
}

\lref\WittenTW{
  E.~Witten,
  ``Global Aspects Of Current Algebra,''
  Nucl.\ Phys.\ B {\bf 223}, 422 (1983).
}

\Title{hep-th/0605206} {\vbox{\centerline{}
\bigskip
\centerline{Axions In String Theory  }}}
\smallskip\centerline{Peter Svr\v{c}ek}
\smallskip\centerline{\it Department of Physics and SLAC, Stanford
University, Stanford CA 94305/94309 USA}
\smallskip\centerline{and}\smallskip
\centerline{Edward Witten}
\smallskip
\centerline{\it Institute For Advanced Study, Princeton NJ 08540 USA}

\medskip
In the context of string theory, axions appear to provide the most
plausible solution of the strong CP problem.  However, as has been
known for a long time, in many string-based models, the axion
coupling parameter $F_a$ is several orders of magnitude higher
than the standard cosmological bounds.  We re-examine this problem
in a variety of  models, showing that $F_a$ is close to the GUT
scale or above in many models that have GUT-like phenomenology, as
well as some that do not.  On the other hand, in some models with
Standard Model gauge fields supported on vanishing cycles, it is
possible for $F_a$ to be well below the GUT scale.

\noindent \Date{May, 2006}

\newsec{Review And Introduction}

The purpose of this paper is to reassess the status of axions in
string theory.  We begin with a review and introduction, after
which, in section 2, we  make some general remarks. The rest of
the paper is devoted to analyzing axions in various string models.
The main conclusion is as it has been in the past: there is some
tension between string models of axions and cosmological bounds,
with the axion coupling parameter in many string models being
larger, and hence the axion frequency smaller, than allowed by the
usual cosmological arguments.  However, we also discuss some
string models, both old ones and new ones, that are compatible
with the usual cosmological bounds.

\nref\poly{A. A. Belavin, A. M. Polyakov, A. S. Schwarz, and
Y. S. Tyupkin, ``Pseudoparticle Solutions Of
The Yang-Mills Eqnations,'' Phys. Lett. {\bf B59} (1975) 85.}%
\nref\thooft{G. 't Hooft, ``Symmetry Breaking Through
Bell-Jackiw Anomalies,'' Phys. Rev. Lett. {\bf 37}
(1976) 8.}%
\nref\jackiwrebbi{R. Jackiw and C. Rebbi, ``Vacuum
Periodicity In A Yang-Mills Quantum Theory,'' Phys.
Rev. Lett. {\bf 37} (1976) 172.}%
\nref\cdg{C. G. Callan, Jr., R. F. Dashen, and D. J. Gross,
``The Structure Of The Gauge Theory Vacuum,''
Phys. Lett. {\bf B63} (1976) 334.}%
Since the early days of QCD instanton physics \refs{\poly - \cdg},
it has been understood that to the action of QCD, it is possible
to add a CP-violating interaction \eqn\mobo{I_\theta =
{\theta\over 32\pi^2}\int d^4x \epsilon^{\mu\nu\alpha\beta}\,{\bf
tr}\,F_{\mu\nu}F_{\alpha\beta}.} Here $\theta$ is a coupling
parameter, and the interaction it multiplies measures the
Yang-Mills instanton number.  Since the instanton number, with
appropriate boundary conditions, is an integer (and more
generally, the instanton number is determined by the boundary
conditions modulo an integer), $\theta$ is an angular parameter.

The strong CP problem \ref\pq{R. D. Peccei and H. R. Quinn, ``CP
Conservation In The Presence Of Pseudoparticles,'' Phys. Rev.
Lett. {\bf 38} (1977) 1440.} is the problem of explaining the
extreme smallness of $\theta$, or more precisely of $\bar\theta$,
the effective $\theta$ after rotating away the phases of the quark
bare masses. From the upper limit  on the electric dipole moment
of the neutron, most recently $|d_n|< 6.3 \times 10^{-26}\,e\,{\rm
cm}$ \ref\harris{P. G. Harris, et. al, ``New Experimental Limit on
the Electric Dipole Moment of the Neutron,'' Phys. Rev. Lett. {\bf
82} (1999) 904.}, one has roughly $|\bar\theta|< 3\times
10^{-10}$. From limits on the electric dipole moment of
$^{199}{\rm Hg}$ \ref\rom{M. V. Romalis, W. C. Griffith, and E. N.
Fortson, ``A New Limit On The Permanent Electric Dipole Moment of
$^{199}{\rm Hg}$,'' Phys. Rev. Lett. {\bf 86} (2001) 2505.}, one
has $|\bar\theta|< 1.5\times 10^{-10}$.   At the factor of two
level, the limits on $\bar\theta$ are subject to some QCD
uncertainties  (for a recent discussion, see \ref\psot{M. Pospelov
and A. Ritz,  ``Theta Induced Electric Dipole Moment Of The
Neutron Via QCD Sum Rules,'' Phys. Rev. Lett. {\bf 83} (1999)
2526.}) and some nuclear and atomic uncertainties in interpreting
the results from $^{199}{\rm Hg}$.

Broadly speaking, three solutions to the strong CP problem have
been proposed:

(1) The up quark mass may vanish.

(2) $\bar\theta$ can relax spontaneously to a  suitably small
value if a new light particle, the axion, exists.

(3) Finally, it might be that CP is a valid symmetry
microscopically, and is spontaneously broken in such a way that
$\bar\theta$ naturally turns out to be small.

\nref\weinberg{S. Weinberg, ``The Problem Of Mass,'' Trans. N.Y.
Acad. Sci. {\bf 38} (1977) 185.} \nref\kapman{D. Kaplan and A.
Manohar, ``Current Mass Ratios Of
The Light Quarks,'' Phys. Rev. Lett. {\bf 56} (1986) 2004.}%
 \nref\bns{T. Banks, Y. Nir, and
N. Seiberg, ``Missing (Up) Mass, Accidental Anomalous Symmetries,
and the Strong CP Problem,''
hep-ph/9403203.}%

\def\QCD{{\rm QCD}}
Concerning the first option, it has been realized since early
instanton studies that if one or more quark bare masses vanishes,
then instanton effects, including instanton contributions to
electric dipole moments, vanish. The conceivably realistic way to
implement this in the real world is to assume that $m_u=0$. Though
this seems at first to be inconsistent with estimates of quark
mass ratios from hadron phenomenology \weinberg, it is not clear
that this is so once one allows \refs{\kapman,\bns} for the fact
that the combination $m_dm_s/\Lambda_{\QCD}$ has the same quantum
numbers as $m_u$. However, recent analyses from lattice gauge
theory claim to take such questions into account and to show that
$m_u\not=0$ \ref\gott{C. Aubin et. al. (MILC Collaboration),
``Light Pseudoscalar Decay Constants, Quark Masses, And Low Energy
Constants {}From Three-Flavor Lattice QCD,'' Phys. Rev. {\bf D70}
(2004) 114501, hep-lat/0407028.}.

\nref\wweinberg{S. Weinberg, ``A New Light Boson?'' Phys. Rev.
Lett. {\bf 40} (1978) 223.}%
\nref\wilczek{F. Wilczek, ``Problem Of Strong P and T Invariance
In The Presence Of Instantons,''
Phys. Rev. Lett. {\bf 40} (1978) 279.}%
\nref\kim{J. E. Kim, ``Weak-Interaction Singlet And Strong CP Invariance,''
Phys. Rev. Lett. {\bf 43} (1979) 103.}%
\nref\invis{M. Dine, W. Fischler, and M. Srednicki, ``A Simple
Solution To The Strong CP Problem With
A Harmless Axion,'' Phys. Lett. {\bf B104} (1981) 199.}%
\nref\sikivie{P. Sikivie, ``Experimental Tests Of The `Invisible'
Axion,'' Phys. Rev. Lett. {\bf 51} (1983) 1415.}%
\nref\clarke{R. Bradley, J. Clarke, D. Kinion, L. J. Rosenberg, K.
van Bibber, S. Matsuki, M. Muck, and P. Sikivie, ``Microwave
Cavity Searches For Dark-Matter Axions,'' Rev. Mod. Phys. {\bf 75}
(2003) 777.}%
The second option originated in the work of Peccei and Quinn \pq,
who postulated that there is  a $U(1)$ symmetry (often called a PQ
symmetry), that is conserved except for gauge anomalies, and acts
by chiral rotations on one or more quarks.  If unbroken, such a
symmetry would imply the vanishing of some quark masses, leading
us back to the solution to the strong CP problem with $m_u=0$. It
was assumed in \pq, however, that the quarks get masses from
coupling to a scalar field that carries the $U(1)$ symmetry and
has an expectation value.  In this case, as was soon noted
\refs{\wweinberg,\wilczek}, the PQ breaking leads to a light spin
zero particle -- the ``axion'' -- that gets mass only from QCD
instanton effects.  It originally was assumed that PQ breaking was
tied to electroweak symmetry breaking, but subsequently models
were constructed \refs{\kim,\invis} in which PQ breaking occurs at
a much higher scale.  This leads to an ``invisible'' axion that
interacts extremely weakly.  Experiments that might detect such an
axion were proposed in \sikivie. For a review of the current
status of searches for axionic dark matter, see \clarke. Recently,
significant limits have also come in searches for axions from the
sun \ref\CAST{K. Zioutas et. al. (CAST collaboration), ``First
Results From The CERN Axion Solar Telescope,'' Phys. Rev. Lett.
{\bf 94} (2005) 121301.}.

\nref\mohapatra{R.~N.~Mohapatra and G.~Senjanovic,
  ``Natural Suppression Of Strong P And T Noninvariance,''
  Phys.\ Lett.\  {\bf B79}, 283 (1978).}%
\nref\nelson{A. Nelson, ``Naturally Weak CP Violation,'' Phys.
Lett. {\bf 136B} (1984) 387.}%
\nref\barr{S. M. Barr, ``Solving The Strong CP Problem Without The
Peccei-Quinn Symmetry,'' Phys. Rev. Lett. {\bf 53} (1984) 329.}%
\nref\schmaltz{G. Hiller and M. Schmalz, ``Solving The Strong CP
Problem With Supersymmetry,'' hep-ph/0105254.}%
\nref\babu{K.~S.~Babu, B.~Dutta and R.~N.~Mohapatra,
  ``Solving the strong CP and the SUSY phase problems with parity
symmetry,''  Phys.\ Rev.\ D {\bf 65}, 016005 (2002).}
 The third idea was developed in
\refs{\mohapatra-\barr}.  There is no problem in beginning with an
underlying CP-invariant model, so that the bare value of $\theta$
vanishes, and then breaking CP spontaneously.  However, the
effective $\theta$ measured at low energies receives a contribution
from the phase of the determinant of the quark mass matrix.  The
quark mass matrix is complex (as we know from studies of CP
violation in weak interactions). The trick in using an underlying CP
symmetry to make $\bar\theta$ small is to ensure that the
determinant of the quark mass matrix is real and positive (to
sufficient accuracy), although that matrix is complex. This is a
little delicate, but can be done in a technically natural, and even
fairly elegant, way \refs{\mohapatra-\babu}.

\nref\kim{J. E. Kim, ``Light Pseudoscalars, Particle Physics, and Cosmology,''
Phys. Reports {\bf 150} (1987) 1.}%
\nref\turner{M. Turner, ``Windows On The Axion,'' Phys. Reports {\bf 197}
(1990) 67.}%
\nref\aa{J. Preskill, M.  Wise, and F. Wilczek, ``Cosmology Of The
Invisible Axion,''
 Phys. Lett. {\bf B120}
(1983) 127.}%
\nref\bb{L. F. Abbott and P. Sikivie, ``A Cosmological Bound On
The Invisible Axion,'' Phys. Lett. {\bf B120}
(1983) 133.}%
\nref\cc{M. Dine and W. Fischler, ``The Not So Harmless Axion,''
 Phys. Lett. {\bf B120} (1983) 137.}%
 In this paper, we consider
primarily the axion hypothesis. As explained in early reviews
\refs{\kim,\turner}, in addition to laboratory constraints if
$F_a$ is close to the weak scale, this hypothesis is subject to a
large variety of astrophysical constraints.  If the axion coupling
parameter $F_a$ (see section 2 for its definition) is less than
about $10^9$ GeV, then the axion couples too strongly and too many
axions are produced in various astrophysical environments, causing
red giants to cool too rapidly, for example. There is also
astrophysical trouble if the axion coupling is too {\it weak}, in
other words if $F_a$ is too large \refs{\aa - \cc}. With standard
cosmological assumptions, if $F_a$ is greater than about $10^{12}$
GeV, the early universe produces too much axionic dark matter
relative to what we see.

Since the upper bound on $F_a$ leads to some tension with string
theory,  we will describe it in more detail. It is assumed that
the early universe starts out at very high temperatures with a
random value of the axion field. It is hard, according to the
argument, for the initial value of the axion field to be anything
but random, since the ``correct'' value of the axion field which
minimizes the energy depends on the phases of the light quark
masses, and these are irrelevant in the very early universe.  (In
addition, inflationary fluctuations might have randomized the
initial value of the axion field.) The axion field has a potential
energy, which at low temperatures is of order
$F_{\pi}^2m_{\pi}^2$, that is determined by QCD effects and is
irrelevant in the very early universe.  (To evaluate the bound on
$F_a$ precisely, one must take into account the temperature
dependence of the effective potential.)  The natural frequency of
oscillation of the axion field is of order $F_{\pi}m_{\pi}/F_a$.
Once the universe cools enough so that this frequency exceeds the
Hubble constant, the axion field begins to oscillate in its
potential.  The oscillations describe a bose-condensed ensemble of
nonrelativistic axion particles, and once the field begins to
oscillate, the number density of axion particles diminishes as the
universe expands.  The larger is $F_a$, the later the oscillations
begin, and the greater is the energy density of axions at the end.
For $F_a$ greater than about $10^{12}$ GeV, the energy density of
axions in today's universe, calculated on these assumptions, would
exceed the dark matter density that we actually observe.  The net
effect of the astrophysical and cosmological bounds is to place
$F_a$ in a range from about $10^9$ to $10^{12}$ GeV -- smaller, as
we will discuss, than is natural in many string models.

An obvious question about the axion hypothesis is how natural it
really is. Why introduce a global PQ ``symmetry'' if it is not
actually a symmetry? What is the sense in constraining a theory so
that the classical Lagrangian possesses a certain symmetry if the
symmetry is actually anomalous?\foot{A similar question can be
asked about the hypothesis that $m_u=0$, as QCD has no additional
symmetries when $m_u=0$. Nevertheless, technically natural models
that lead to a solution of the strong CP problem via $m_u=0$ do
exist.  For an example, see \bns.  Incidentally, $m_u$ can be
rigorously measured, in principle, from certain OPE coefficients
that violate the chiral symmetry of up quarks (violation of the
same symmetry by instantons is softer at short distances); an
example of a precise definition of $m_u$  is as follows. Take the
current operator $J= \bar u \gamma_\mu u$ and let $S$ be the
scalar operator $S= \bar u u.$ In the operator product expansion
$J(x)J(0) \sim ... + f(x) S(0) + ...$, the operator $S$ appears
with a nonzero coefficient function $f(x)$ whether $m_u$ is zero
or not. One can define $m_u$ as the coefficient of $1/|x|^2$ in
$f(x)$, or more precisely, as the coefficient of $(-\ln
|x|)^a/|x|^2,$ where the exponent $a$ comes from the usual
one-loop anomalous dimension of the operator $S$. If $m_u=0$, then
$f(x)$ vanishes in perturbation theory (but receives instanton
contributions) and is less singular as $x$ goes to zero than
$(-\ln |x|)^a/|x|^2$.}

 It could be argued that the best evidence that PQ
``symmetries'' are natural comes from string theory, which
produces them without any contrivance. Soon after the discovery of
the Green-Schwarz anomaly cancellation mechanism \ref\gs{M. B.
Green and J. H. Schwarz, ``Anomaly Cancellation In Supersymmetric
$D=10$ Gaulsoge Theory And Superstring Theory,'' Phys. Lett. {\bf
B149} (1984) 117.} expanded the possible scope of string
phenomenology, it was recognized \ref\ewitten{E. Witten, ``Some
Properties of $O(32)$ Superstrings,'' Phys. Lett. {\bf B149}
(1984) 351.} that the terms in the low energy effective action
that lead to  anomaly cancellation  also cause certain light
string modes to behave as axions (for reviews, see chapter 14.3.2
of \ref\gsw{M. B. Green, J. H. Schwarz, and E. Witten, {\it
Superstring Theory}, vol. 2 (Cambridge University Press, 1987).},
or section 7.6 of \kim). In certain cases, some of the would-be
axions also get masses from a Higgs mechanism, and one is left
over with a global PQ symmetry. These statements can be unified by
saying that the string compactifications always generate PQ
symmetries, often spontaneously broken at the string scale and
producing axions, but sometimes unbroken.

This convincingly shows that axions and PQ symmetries are natural,
but in the original models, the resulting axion mass parameter
$F_a$ was too large for the cosmological bounds. This problem
motivated a number of early contributions. In \ref\barr{S. Barr,
``Harmless Axions In Superstring Theories,'' Phys. Lett. {\bf
B158} (1985) 397.}, it was proposed to circumvent the problem by
taking an {\it unbroken} PQ symmetry from the string and
spontaneously breaking it at lower energies, producing an axion of
smaller $F_a$ (but less simply described as a string mode than the
axions found in \ewitten). In \ref\linde{A. Linde, ``Inflation And
Axion Cosmology,'' Phys. Lett. {\bf B201} (1988) 437.}, it was
proposed that, for anthropic reasons, the usual cosmological bound
on $F_a$ might be invalid, avoiding the contradiction of this
bound with most of the string models. The clash between the
cosmological bound on $F_a$ and the simplest string theory
predictions for it was also emphasized in the review article \kim.

More recent developments have led to the emergence of many
possible new string-based models of particle physics.  In this
paper, we will reassess the string theory predictions for $F_a$ in
the light of these new developments. We first reconsider the
heterotic string.  The value of $F_a$ for the
``model-independent'' axion of the weakly coupled heterotic string
has been computed in \ref\choi{K. Choi and J. E. Kim, ``Harmful
Axions In Superstring MOdels,'' Phys. Lett. {\bf B154} (1985) 393,
``Compactification And Axions In $E_8\times E_8'$ Superstring
Models,'' Phys. Lett. {\bf B165} (1985) 689.} and more precisely
recently in  \ref\thomas{P. Fox, A. Pierce, and S. Thomas,
``Probing A QCD String Axion With Precision Cosmological
Measurements,'' hep-th/0409059.}.  The value is $F_a=1.1\times
10^{16}$ GeV if one assumes the usual running of $\alpha_s$ up to
the string scale.  We work out analogous predictions for the
``model-dependent'' heterotic string axions. We then go on to
consider other models, such as strongly coupled $E_8\times E_8$
heterotic strings (axions were previously discussed in these
models in \ref\bd{T. Banks and M. Dine, ``Couplings And Scales In
Strongly Coupled Heterotic String Theory,'' Nucl. Phys. {\bf B479}
(1996) 173.}), $M$-theory on a manifold of $G_2$ holonomy, and
models based on $D$-branes and orientifolds.

A general conclusion is that, in many models, it is difficult to
push $F_a$ drastically below $1.1\times 10^{16}$ GeV and easier to
increase it closer to the reduced Planck mass, $\sqrt{1\over 8\pi
G_N}\sim 2.4 \times 10^{18}$ GeV. (It has been argued earlier that
it is hard to make $F_a$ -- or its analog for other light spin zero
modes -- much {\it higher} than the Planck scale \ref\bmfe{T. Banks,
M. Dine, P. J. Fox, and E. Gorbatov, ``On The Possibility Of Large
Axion Decay Constants,'' JCAP 0306:001 (2003), hep-th/0303252.}. We
do not have anything new to say about this.) Examples include most
models with GUT-like phenomenology, and also models in which QCD
gauge fields are supported on $D3$-branes.  On the other hand, there
are also models in which $F_a$ can be lower. One early approach
\barr\ to lowering $F_a$ below the GUT scale remains potentially
valid from a modern point of view and can be compatible with GUT
phenomenology, though possibly not with low energy supersymmetry. If
one is willing to abandon GUT phenomenology, there is another
possibility, in which QCD gauge fields are supported on a
``vanishing cycle.''  We discuss various examples, beginning in
section 4.  These models have string or Kaluza-Klein scales well
below the Planck scale, and hence their very early cosmology may be
exotic.  Related ideas have been discussed in \ConlonTQ\ (which
appeared while the present paper was in gestation).

{}From a modern point of view, PQ symmetries generated from string
theory are always explicitly violated by instantons of some kind
(in addition to the low energy QCD instantons); candidates include
worldsheet instantons
 \ref\dsww{M. Dine, N. Seiberg, X.-G. Wen, and E.
 Witten, ``Nonperturbative Effects On The String World Sheet,''
 Nucl. Phys. {\bf B278} (1986) 769.}, brane instantons
 \ref\strom{K. Becker, M. Becker, and A. Strominger, ``Five-Branes, Membranes,
 And Nonperturbative String Theory,'' Nucl. Phys. {\bf B456} (1995) 130,
 hep-th/9507158.}, gauge instantons from other factors of the gauge group,
 and gravitational instantons.  Such effects, of course,
 can be exponentially small.
 One of the necessary conditions for solving the
 strong CP problem via string-derived instantons is that explicit
 breaking of the relevant PQ symmetry by non-QCD effects must be
 much smaller (at least $10^{10}$ times smaller, to make $\bar\theta$ small enough)
 than breaking due to QCD instantons.  Thus, in the
 various models, in addition to estimating $F_a$, we also estimate
 the actions of the relevant instantons.  Suppressing the
 instantons is a significant constraint on models and in many
 cases favors having supersymmetry survive at least somewhat below
 the GUT scale.  (This is the main reason that supersymmetry
 enters our discussion. Though for convenience we consider
 compactification manifolds that preserve supersymmetry, this will not play
 an essential role in most of the analysis.  The Barr mechanism \barr\ may be an exception.)

What about other solutions to the strong CP problem in string
theory?  If we leave aside lattice gauge theory evidence \gott\
that $m_u\not= 0$, one could try to embed $m_u=0$ in string theory
by leaving a string-based PQ symmetry {\it unbroken}.  The hard
part would probably be to ensure  that the PQ symmetry enforces
$m_u=0$ without setting to zero any other quark or charged lepton
masses. As for the third approach of explaining the strong CP
problem by assuming that CP is spontaneously broken in a way that
leaves the determinant of the quark mass matrix real and positive,
string theory passes the first hurdle in that in almost all string
theory compactifications, there is a locus in moduli space at
which CP is unbroken, and thus CP can be interpreted as a
spontaneously broken symmetry. (For example, see section 16.5.1 in
\gsw.) This still leaves much ground to cover, and we do not know
if a mechanism such as that of \refs{\nelson,\barr} can be
embedded in string theory.

As we have already noted, many string models give $F_a$ too large
for the standard cosmological bounds, and in fact the value
$F_a=1.1\times 10^{16}$ GeV first given in \refs{\choi,\thomas},
or something relatively near it, arises in a number of different
models. This value corresponds to an axion frequency close to 130
KHz. Most models we consider lead to $F_a$ in the range from
roughly $10^{15}$ GeV to the reduced Planck mass. At the reduced
Planck mass, the axion frequency is a little less than 1 KHz.

While these values clash with standard cosmological reasoning,
proposals have been made that would relax the cosmological bounds.
One proposal involves anthropic arguments \linde; others involve
arranging so that QCD was actually strongly coupled in the early
universe \ref\dvali{G. Dvali, ``Removing The Cosmological Bound On
The Axion Scale,'' hep-ph/9505253.}, late entropy production due
to particle decays \ref\kaw{M. Kawasaki, T. Moroi, and T.
Yanagida, ``Can Decaying Particles Raise The Upper Bound On The
Peccei-Quinn Scale?'' Phys.Lett. {\bf B383} (1996) 313,
hep-ph/9510461}, or drastically reducing the energy scale of
inflation \nref\bd{T. Banks and M. Dine, ``The Cosmology Of String
Theoretic Axions,''
Nucl. Phys. {\bf B505} (1997) 445, hep-th/9608197.}%
 \nref\dine{T. Banks, M. Dine, and M. Graesser, ``Supersymmetry, Axions, And
Cosmology,'' Phys. Rev. {\bf D68} (2003) 075011, hep-ph/0210256.}%
\refs{\bd,\dine}. A variant of the last proposal (stressed to us
by P. Steinhardt)  is the cyclic model of the universe
\ref\sttu{P. Steinhardt and N. Turok, ``A Cyclic Model Of The
Universe,'' Science {\bf 296} (2002) 1436, ``The Cyclic Model
Simplified,'' New Astron. Rev. {\bf 49} (2005) 43,
astro-ph/0404480.}, in which the universe might never reach the
high temperatures needed to create an excess of axionic dark
matter. The anthropic proposal  would lead one to expect that
axionic dark matter would be significant in the universe, a point
made in \linde\ and reconsidered recently \ref\wwilczek{F.
Wilczek, ``A Model Of Anthropic Reasoning, Addressing The Dark To
Ordinary Matter Coincidence,'' to appear in {\it Universe Or
Multiverse?}, ed. B. Carr, hep-ph/0408167.}.  A model with $F_a$
large also has to survive certain other cosmological constraints
\thomas. At any rate, an experiment capable of finding or
excluding axionic dark matter in the most relevant frequency range
-- from about 1 KHz to a few MHz -- would greatly clarify things.
It has been suggested \nref\romalis{M.
Romalis, private communication.}%
\nref\tthomas{S. Thomas, private communication.}%
\refs{\romalis,\tthomas} that this might be accomplished with an
experiment using LC circuits.

\newsec{General Remarks }

The QCD instanton number can be written in many equivalent ways.
It is \eqn\instnumber{N = {1\over 32\pi^2}\int d^4x
\epsilon^{\mu\nu\alpha\beta}\,{\bf tr}\,F_{\mu\nu}F_{\alpha\beta}
={1\over 16 \pi^2}\int d^4x \,{\bf tr} \,F_{\mu\nu}\tilde
F^{\mu\nu},} where $\tilde
F_{\mu\nu}=\half\epsilon_{\mu\nu\alpha\beta}F^{\alpha\beta}$. We
use the sign convention $\epsilon^{0123}=1.$ The trace is taken in
the three-dimensional representation of $SU(3)$. Alternatively, if
we introduce the two-form $F=\half F_{\mu\nu}dx^\mu\wedge dx^\nu$,
then \eqn\noggo{N={1\over 8\pi^2}\int \,{\bf tr}\, F\wedge F.}
Finally, if we write $F_{\mu\nu}=F_{\mu\nu}^aT_a$, where $T_a$ is
a basis of the Lie algebra, normalized to ${\bf tr}\,T_aT_b=\half
\delta_{ab}$, then \eqn\lolop{N={1\over 64\pi^2}\int d^4x
\epsilon^{\mu\nu\alpha\beta} F_{\mu\nu}^aF_{\alpha\beta}^a.} In
writing the instanton number, we normalize the gauge fields so
that the covariant derivative $D_\mu=\partial_\mu+iA_\mu$ is
independent of the gauge coupling $g$, which instead appears in
the gauge kinetic energy \eqn\ubuc{-{1\over 2g^2}\int d^4x\, {\bf
tr}\, F_{\mu\nu}F^{\mu\nu}=-{1\over 4g^2}\int d^4x\,
F_{\mu\nu}^aF^{\mu\nu\,a}.} (We write the action in Lorentz metric
with signature $-+++$.)  It follows that the action of an
instanton, that is a field with $F=\tilde F$ and $N=1$, is
\eqn\instac{I={8\pi^2\over g^2}={2\pi\over\alpha_s},} where
$\alpha_s=g^2/4\pi$.

The axion is a spin zero field $a$ with a PQ shift symmetry $a\to
a+{\rm constant}$ that is broken only, or primarily, by the
effects of QCD instantons.  The coupling of $a$ to QCD instantons
is \eqn\nono{{r\over 32\pi^2}\int d^4x\,a\,
\epsilon^{\mu\nu\alpha\beta}\,{\bf
tr}\,F_{\mu\nu}F_{\alpha\beta},} where $r$ is a constant. In all
of the string and $M$-theory models, the axion turns out to be a
periodic variable; we normalize it so that the period is $2\pi$.
As $N$ is an integer, the action is well-defined mod $2\pi$ if $r$
is an integer, as is the case in the string models.

The axion field also has a kinetic energy, $\half F_a^2\partial_\mu
a\partial^\mu a$, for some constant $F_a$. The low energy effective
action of the axion thus includes the terms \eqn\jumbo{\Delta I=\int
d^4x\left(-{F_a^2\over 2}\partial_\mu a\partial^\mu a+ {ra\over
32\pi^2} \epsilon^{\mu\nu\alpha\beta}\,{\bf
tr}\,F_{\mu\nu}F_{\alpha\beta}\right).} $F_a$ is called the axion
decay constant or coupling parameter, because axion couplings are
proportional to $1/F_a$.  For example, if we introduce a rescaled
axion field $\tilde a = F_a a$, then the kinetic energy becomes
canonical, and the axion coupling is proportional to $1/F_a$:
\eqn\umbo{\Delta I=\int d^4x\left(-{1\over 2}\partial_\mu\tilde
a\partial^\mu \tilde a+ {r\tilde a\over 32\pi^2 F_a}
\epsilon^{\mu\nu\alpha\beta}\,{\bf
tr}\,F_{\mu\nu}F_{\alpha\beta}\right).} As explained in the
introduction, the main result of the present paper is that in a wide
range of string models, $F_a$ is in the range from roughly the GUT
scale to the reduced Planck scale, and thus is above the range
favored by the usual cosmological bounds.

To see that existence of an axion solves the strong CP problem,
first note that if an axion is present, the physics is independent
of the QCD vacuum angle $\theta$.  The $\theta$-dependence of the
action is an additional term, \eqn\novo{{\theta\over 32\pi^2}\int
d^4x \epsilon^{\mu\nu\alpha\beta}\,{\bf
tr}\,F_{\mu\nu}F_{\alpha\beta},} and if an axion field $a$ is
present, this term can be eliminated by shifting $a$ by a
constant, $a\to a-\theta/r$. In effect, the existence of an axion
promotes $\theta$ to a dynamical field $ra$; the vacuum
expectation value of $a$ must be determined to minimize the
energy.

To compute the potential energy as a function of $a$, one must know
how to calculate the vacuum energy of QCD as a function of $\theta$.
Because the up and down quark masses are so small, this can
conveniently be done using current algebra, as reviewed in
\ref\weinbergbook{S. Weinberg,  {\it The Quantum Theory Of Fields},
Vol. II (Cambridge University Press, 1996).}, section 23.6. In the
two flavor case (the $u$ and $d$ quarks are so much lighter than the
$s$ quark that the latter can be neglected), we describe low energy
pion physics by an $SU(2)$-valued field $U$ with an effective
Lagrangian \eqn\hingo{ L=-{F_\pi^2\over 16}\,\,{\bf tr}
\,\partial_\mu U\partial^\mu U^{-1} +{v\over 2} \,{\bf tr}\,\left( M
U+\bar M U^{-1}\right).} Here, $ M$ is the quark mass matrix, which
at $\theta=0$ we can take to be $M=\left(\matrix{m_u & 0 \cr 0 &
m_d\cr}\right)$, with $m_u,$ $m_d$ being real and positive; also, to
get the right pion mass and couplings, $F_\pi = 184$ MeV, and
$v(m_u+m_d)= F_\pi^2m_\pi^2/4$. One can include $\theta$ by
replacing (for example) $m_u\to e^{i\theta}m_u$.  So, upon promoting
$\theta$ to a field $ra$, we take \eqn\uvcu{M=\left(\matrix{
m_ue^{ira} &  0 \cr 0 & m_d\cr}\right).} The effective potential for
the light field $a$ is obtained by minimizing the potential
$V(U,a)={v\over 2}{\bf tr}\,(MU+\overline M U^{-1})$ as a function
of $U$ for fixed $a$. The minimum energy is at $a=0$; the term
quadratic in $a$ turns out to be \eqn\gofo{V(a)={a^2\over 8} r^2
F_\pi^2m_\pi^2 {m_um_d\over (m_u+m_d)^2}.} So allowing for the
normalization of the axion kinetic energy $\half \int d^4x
F_a^2(\partial_\mu a)^2$, the axion mass is \eqn\toffo{m_a= {rF_\pi
m_\pi\over 2 F_a}{\sqrt{m_um_d}\over m_u+m_d}.} With the estimate
\weinberg\ that $m_u/m_d\cong 1/1.8$, one has \eqn\goffo{m_a\cong
5.4\times 10^{-10}\,{\rm eV}\cdot {1.1\times 10^{16}\,{\rm GeV}\over
F_a/r}.} The angular frequency of axion oscillations is
$\omega=m_ac^2/\hbar$, and the ordinary frequency is
\eqn\tono{\nu={\omega\over 2\pi}= 130\,\,{\rm KHz}\cdot {1.1\times
10^{16}\,{\rm GeV}\over F_a/r}.} We will see that in many string
models, $F_a/r$ is naturally fairly near $1.1\times 10^{16}$ GeV,
the value found \refs{\choi,\thomas} for the model-independent axion
of perturbative heterotic strings.

In string models, there are inevitably, apart from low energy QCD
instantons, some other instantons -- such as worldsheet or
membrane instantons -- that violate the PQ symmetry.  Suppose that
the axion couples to a stringy instanton of action $S_{inst}$ with
a natural mass scale $M$. If there is no suppression due to
supersymmetry, the instanton will generate a potential of the
general form $-M^4\exp(-S_{inst})\exp(i(a+\psi))$ for some phase
$\psi$. Since we defined the axion field so that the low energy
QCD contribution to the axion potential is minimized at $a=0$,
$\psi$ really arises from a mismatch in phase between the high
scale instantons and the analogous phase in the low energy
contribution, which is affected by things such as the light quark
masses. Lacking a special theory of the light quark masses, we
interpret $\psi$ as an arbitrary phase that is not likely to be
particularly small.  The axion potential induced by such
instantons and anti-instantons is of order  \eqn\ubu{\tilde V(a)
=- 2M^4\exp(-S_{inst})\cos(a+\psi).}  Upon minimizing $V(a)+\tilde
V(a)$, we get in order of magnitude \eqn\ubu{r^2a\sim
M^4\exp(-S_{inst})/F_\pi^2m_\pi^2.} Since $ra$ is the effective
QCD theta angle, we require $|ra|<10^{-10}$ to agree with
experimental constraints, so we need
\eqn\nubu{\exp(-{S_{inst}})<10^{-10}r{F_\pi^2m_\pi^2\over M^4}.}
This is a severe constraint.  For example, we if set $M$ to the
reduced Planck mass $2.4 \times 10^{18}$ GeV, we need
\eqn\gogoo{S_{inst}>200.} This is a significant constraint on
models, as has been stressed in \bd\ (where it was argued that
this constraint favors the strongly coupled heterotic string over
the perturbative heterotic string) and as we will discuss further.
Many models with large extra dimensions and a reduced string scale
do not help much in reducing $F_a$, as we will see. But by
reducing the mass $M$ in \nubu, they do relax the requirements for
the instanton action.

Those requirements are also relaxed if supersymmetry survives to a
scale lower than $M$.  In this case, instead of generating a
potential proportional to the real part of
$M^4\exp(-S_{inst}+i(a+\psi))$, the instanton may generate a
contribution to the superpotential proportional to
$W_0=M^3\exp(-S_{inst}+i(a+\psi))$.  When we evaluate the ordinary
potential $V=|DW/D\Phi|^2-3G_N|W|^2$, if we simply set $W=W_0$,
the $a$-dependence will cancel.  An $a$-dependence of $V$ can
arise, for example, from an interference of the one-instanton term
with a contribution to $W$ from some other source.  We suppose
that this other contribution breaks supersymmetry at some scale
$\mu$, with $DW/D\Phi\sim \mu^2$. Then the ordinary potential
contains terms that are roughly
$M^2\mu^2\exp(-S_{inst})\cos(a+\psi)$, and \nubu\ is replaced by
\eqn\ubu{\exp(-S_{inst})<10^{-10}{F_\pi^2m_\pi^2\over M^2\mu^2}.}
If $\mu$ is even a few orders of magnitude below $M$, or $M$ below
the reduced Planck mass, this  makes life easier for many models.
Actually, supersymmetry can lead to more suppression of high scale
instanton effects than we have just described if the instantons
have fermion zero modes beyond those required by supersymmetry and
contribute not to the superpotential but to higher order chiral
operators, as considered in \ref\beas{C. Beasley and E. Witten,
``New Instanton Effects In Supersymmetric QCD,''  JHEP {\bf
0501:056} (2005), hep-th/0409149.}.

One important point is that the instanton that gives the dominant
contribution to the axion mass may not entirely break the shift
symmetry of the axion.  It may reduce this symmetry to an $n$-fold
discrete symmetry, for some integer $n$. For example, for an axion
that actually solves the strong CP problem, the dominant instanton
is an ordinary QCD instanton, and the integer $n$ coincides with
the integer $r$ of the axion coupling in eqn. \nono. The shift
symmetries that are not explicitly broken by the instanton are
spontaneously broken by the axion expectation value.  That will
result, in this approximation, in having $n$ degenerate vacua
differing by the value of the axion field. However, in the context
of string theory, one expects that this degeneracy will always be
lifted at a lower energy by some other, subdominant, type of
instanton. The full collection of string theory instantons
(including worldsheet instantons, gauge and gravitational
instantons, wrapped branes, etc.) is expected to fully break the
axion symmetries. This statement roughly means that the theory has
the full set of branes and other topological defects allowed by
the periodicity of the axions, just as \ref\jplo{J. Polchinski,
``Monopoles, Duality, And String Theory,'' Int. J. Mod. Phys. {\bf
A19S1} (2004) 145, arXiv:hep-th/0304042.} it has the full set of
magnetic charges allowed by Dirac quantization and the values of
electric charges.

Apart from its coupling to the QCD instanton density, the axion
may have additional couplings to Standard Model fields. The PQ
shift symmetry allows the axion to have arbitrary derivative
couplings to quarks and leptons. In addition, it may have
couplings just analogous to \nono\ to the ``instanton densities''
of other gauge fields and gravity. The coupling to
electromagnetism is of particular importance, since it is the
basis for axion searches \refs{\sikivie-\CAST}. For example, in an
$SU(5)$ grand unified theory, the electromagnetic field
$f_{\mu\nu}$ appears in the underlying $SU(5)$ gauge theory via an
ansatz \eqn\toggo{ F_{\mu\nu}=f_{\mu\nu}\left(\matrix{ -1/3 & & &&
\cr
                                        & -1/3 & & & \cr
                                        &      & -1/3 & & \cr
                                        &       &      & 1 & \cr
                                        & & & &~ 0 \cr}\right).}
If we extend the trace in \umbo\ to a trace over the fundamental
representation of $SU(5)$, we get ${\bf
tr}\,F_{\mu\nu}F_{\alpha\beta}=(4/3)f_{\mu\nu}f_{\alpha\beta}+\dots$,
so in an $SU(5)$ grand unified theory, the coupling of the axion
to electromagnetism is \eqn\corko{{4r\over 3}{1\over 32\pi^2
F_a}\int d^4x \,\tilde
a\epsilon^{\mu\nu\alpha\beta}\,f_{\mu\nu}f_{\alpha\beta}.} It is
convenient to rewrite this in terms of a canonically normalized
electromagnetic field. In our convention,  as in \ubuc, the gauge
coupling appears as a constant multiplying the action, and the
electromagnetic action is $\int d^4x f_{\mu\nu}f^{\mu\nu}/4e^2$,
with $e$ the charge of the electron.  In terms of the
conventionally normalized electromagnetic tensor $F_{\mu\nu}^{{\rm
em}}=f_{\mu\nu}/e$ with kinetic energy $(F_{\mu\nu}^{{\rm
em}})^2/4$, the coupling becomes
 \eqn\corko{{4r\over 3}{e^2\over 32\pi^2 F_a}\int d^4x \,\tilde
a\epsilon^{\mu\nu\alpha\beta}\,F^{{\rm em}}_{\mu\nu}F^{{\rm
em}}_{\alpha\beta}= -{4r\alpha\over 3\pi F_a}\int d^4x\, \tilde a
\vec E\cdot \vec B.} Here $E_i=F_{0i}^{{\rm em}}$ and
$B_i=\half\epsilon_{ijk}F^{{\rm em}}_{jk}$ are the usual electric
and magnetic fields, and $\alpha=e^2/4\pi\hbar c$. This formula
needs to be corrected, however, as we explain in section 9, because
of mixing between the axion and the $\pi^0$ meson.

In deriving the specific value of the coupling in \corko, we
considered $SU(5)$-like GUT's, but a similar $\tilde a \vec E\cdot
\vec B$ coupling (possibly with a slightly different coefficient)
arises in virtually all string-based models that solve the strong
CP problem via an axion. For example, in heterotic string models,
the axion couplings to strong, weak, and electromagnetic instanton
densities only depend on the ``levels'' of the current algebra (or
the embedding in $E_8$), so with a level 1 embedding of the
Standard Model, the formula \corko\ will hold (with $r=1$), even
if there is no four-dimensional or even ten-dimensional field
theoretic unification. (For gauge coupling constants, the
analogous point is made in \ref\bdfm{T. Banks, L. J. Dixon, D.
Friedan, and E. J. Martinec,  ``Phenomenology And Conformal Field
Theory, Or Can String Theory Predict The Weak Mixing Angle?''
Nucl. Phys. {\bf B299} (1988) 613.}.) The same formula holds, most
commonly with $r=1$, in other string-derived models that
incorporate the usual GUT relations for gauge coupling
unification, even if there is no four-dimensional GUT.  In models
that are not unified, the couplings may be somewhat different.

\bigskip\noindent{\it General Approach}

Each string or $M$-theory model we consider is characterized by an
asymptotic expansion parameter -- $g_s$ in the case of string
theory, $1/R$  in the case of $M$-theory on a $G_2$ manifold of
radius $R$, and so on.  The asymptotic expansion is valid when the
relevant parameter is small, and we expect it to give the correct
order of magnitude when the parameter is of order one.  To assess
how small $F_a$ can be, we calculate in the region where the
expansion parameter is small and then extrapolate to the region
where $g_s$, $1/R$, etc., is of order 1.  The region with (say)
$g_s>>1$ is not accessible in this way, but hopefully is
accessible via a dual asymptotic expansion valid for large $g_s$.
Hence, to the extent that the asymptotic regions that we consider
are representative, we should get a good overview of axion
phenomenology, though in the interior of moduli space our
computations are only valid qualitatively.

\bigskip\noindent{\it Conventions}

In our string theory computations, we will adopt a few conventions
that are aimed to minimize the factors of $2\pi$ in this paper. We
define the string scale by\foot{Apart from the fact that it
eliminates many factors of $2\pi$ from the formulas, a piece of
evidence that this is a good definition is the following.  In
compactification on a circle, the self-dual {\it radius} is
$R=\sqrt{\alpha'}$, and the self-dual {\it circumference} is hence
$L=2\pi \sqrt{\alpha'}=\ell_s$.  The circumference -- the length
of a nontrivial closed geodesic -- is an intrinsic measure of the
size of the circle, while the radius is only a natural notion if
the circle is embedded in a plane.} $\ell_s=2\pi\sqrt{\alpha'}$,
and we normalize $p$-form fields to have integer periods. Thus,
our NS $B$-field is related to the usual one $B_{conv}$ by
$B_{conv}=\ell_s^2B$.  As a result, the coupling of the $B$-field
to the string worldsheet $\Sigma$, usually
$(i/2\pi\alpha')\int_\Sigma B_{conv}$, becomes $2\pi i\int_\Sigma
B$. The contribution to the action proportional to the area of the
worldsheet becomes $(2\pi/\ell_s^2)\int d^2\sigma\sqrt{\det G}$.

We follow \ref\polchinski{J. Polchinski, {\it String Theory},
Cambridge University Press (Cambridge, 1998).}, eqn. (13.3.22), in
normalizing the Type II dilaton $\phi_{II}$ and Type II string
coupling $g_{II}=\exp(\phi_{II})$ so that the ten-dimensional
gravitational coupling $\kappa$ (appearing in the Einstein action
$\int d^{10}x\sqrt g R/2\kappa^2$) is \eqn\molmo{\kappa^2={1\over
2} (2\pi)^7g_{II}^2(\alpha')^4={g_{II}^2\ell_s^8\over 4\pi}.} This
convention ensures that $SL(2,\Bbb{Z})$ duality acts by $g_{II}\to
1/g_{II}$, with no numerical factor.  We will use the same
convention \molmo\ for Type I and heterotic superstring theories
(with, of course, $g_{II}$ replaced by the corresponding string
couplings $g_I=\exp(\phi_I)$ and $g_h=\exp(\phi_h)$).  This
ensures that under heterotic-Type I duality, we have $g_I=1/g_h$
with no numerical constant.  On the other hand, one must be
careful of a factor of two in duality between Type I on a torus
$\Bbb{T}^n$ and Type II on the orientifold $\Bbb{T}^n/\Bbb{Z}_2$.
Note also the definition $\kappa_{10}^2=\kappa^2/g^2$ in
\polchinski, used for all of the string theories,

With our definition $\ell_s=2\pi\sqrt{\alpha'}$, the tension of a
$Dp$-brane in Type II superstring theory becomes (following
\polchinski, eqn. (13.2.3)) \eqn\noggo{\tau_p={2\pi\over
g_{II}\ell_s^{p+1}}.}

Similarly, to minimize factors of $2\pi$ in eleven-dimensional
supergravity, we introduce an $M$-theory length $\ell_{11}$ by
$\ell_{11}=2\pi/M_{11}$, where $M_{11}$ is defined in terms of
Type IIA superstring parameters by eqn. (14.4.6) of \polchinski.
In these units, the membrane and fivebrane tensions (in view of
\polchinski, eqns. (14.4.11) and (14.4.18)) are
\eqn\polygo{\tau_{M2}={2\pi\over
\ell_{11}^3},~~\tau_{M5}={2\pi\over \ell_{11}^6}.} Moreover, from
eqn. (14.4.5) of that reference, we have
$2\kappa_{11}^2=\ell_{11}^9/2\pi$. Moreover, we normalize the
three-form field $C$ to have integer periods; this means that our
$C$ is related to the analogous field $C'$ used in \polchinski\ by
$C= C'/\ell_{11}^3$.  The action of eleven-dimensional
supergravity, from eqn. (12.1.1), becomes
\eqn\olygo{S_{11}=2\pi\int \left({1\over
\ell_{11}^9}d^{11}x\sqrt{-g}R-{1\over 2\ell_{11}^3}G\wedge \star G
-{1\over 6}C\wedge G\wedge G\right).} The Hodge star operator
$\star$ is defined so that if $W$ is a $p$-form, then $W\wedge
\star W = {\sqrt{-g}\over
p!}W_{\mu_1\mu_2\dots\mu_p}W^{\mu_1\mu_2\dots\mu_p}$.

\def\tr{{\rm tr}}\def\Tr{{\rm Tr}}%
We write ${\bf tr}$ for the trace in the fundamental
representation of $SU(N)$, $\tr$ for the trace in the fundamental
representation of $SO(M)$, and $\Tr$ for the trace in the adjoint
representation of $E_8$  or $SO(32)$.  If $SU(N)$ is embedded in
$SO(2N)$, then $\tr=2{\bf tr}$. For $SO(32)$, one has
$\tr=\,\Tr/30$, and for $E_8$ one defines $\tr=\Tr/30$.  When we
pick a basis of the Lie algebra, we require
$\tr\,t^at^b=\delta^{ab}$ and hence (for $SU(N)$) ${\bf
tr}\,t^at^b=\delta^{ab}/2$.

In $M$-theory on a manifold with a boundary $ M^{10}$, $E_8$ gauge
fields appear on the boundary.  The gauge action is
\eqn\olfog{S_{Y\negthinspace M}=-{1\over
4(2\pi)\ell_{11}^6}\int_{M^{10}} \tr F\wedge \star F.}

\newsec{Axions In Weakly Coupled Heterotic String Theory}

The relevant part of the ten-dimensional low energy Lagrangian of
the heterotic string is, from eqn. (12.1.39) of \polchinski,
\eqn\hete{\eqalign{L&={1\over 2 \kappa_{10}^2} \sqrt{-g} R- {1\over
4\kappa_{10}^2} H\wedge \star H -{\alpha'\over 8\kappa^2_{10}} \tr
F\wedge \star F \cr &= {2\pi\over g_s^2\ell_s^8}
\sqrt{-g}R-{2\pi\over g_s^2\ell_s^4}\half H\wedge\star H -{1\over
4(2\pi) g_s^2\ell_s^6}\tr F\wedge \star F.}} Here $R$ is the Ricci
scalar, $H$ the field strength of the two-form field $B$, and $F$
the $E_8\times E_8$ or $SO(32)$ curvature; we have used conventions
explained at the end of the section 2 together with the relation
$\kappa_{10}^2/g_{10}^2=\alpha'/4$ from eqn. (12.3.36) of
\polchinski.

To reduce to four dimensions, we compactify on a six-manifold $Z$
(not necessarily Calabi-Yau) with volume $V_Z$.  The
four-dimensional spacetime (which might be Minkowski spacetime) we
call $M$. The relevant terms in the four-dimensional effective
action include \eqn\hetef{S= {M_P^2\over 2}\int d^4x(-g)^{1/2}
R-{1\over 4 g_{Y\negthinspace M}^2} \int
d^4x\sqrt{-g}\tr\,F_{\mu\nu}F^{\mu\nu}- {2\pi V_Z \over g_s^2
l_s^4} \int \left({1\over 2} H\wedge \star H\right),} where the
four-dimensional reduced Planck mass $M_P$ and Yang-Mills coupling
$g_{Y\negthinspace M}$ are \eqn\fdc{ M_P^2= 4\pi {V_Z\over
g_s^2\ell_s^8}} and \eqn\fcq{g^2_{Y\negthinspace M}=4\pi
{g_s^2\ell_s^6\over V_Z}.} So $\alpha_{Y\negthinspace
M}=g_{Y\negthinspace M}^2/4\pi$ is given by
\eqn\apl{\alpha_{Y\negthinspace M}={g_s^2\ell_s^6\over V_Z}.}

If we assume the usual level one embedding of the Standard Model
gauge fields in $E_8\times E_8$, then we can identify
$\alpha_{Y\negthinspace M}$ with $\alpha_{GUT}$, the unified gauge
coupling at the string (or ``GUT'') scale. More generally, if we
make a level $k$ embedding of the Standard Model in the heterotic
string (for discussion, see \ref\ib{L. Ibanez, ``Gauge Coupling
Unification: Strings Versus SUSY Guts,'' Phys. Lett. {\bf B318}
(1993) 73, hep-ph/9308365.}), then $\tr\,F_{\mu\nu}F^{\mu\nu}$,
when evaluated for Standard Model gauge fields, has an extra
factor of $k$, which reduces $\alpha_{GUT}$ by a factor of $k$.
Thus we have in general
\eqn\mormo{\alpha_{GUT}={\alpha_{Y\negthinspace M}\over
k}={g_s^2\ell_s^6\over kV_Z}.}

In general, it may be that not all factors in the Standard Model
gauge theory have the same $k$. To determine the axion mass and
frequency, we really need the coupling of the axion to QCD.  So we
let $k$ denote the level of the color $SU(3)$ embedding in
$E_8\times E_8$.  In all models considered in this paper, we
define $\alpha_C$ to be the strong coupling constant $\alpha_s$
evaluated at the string scale or compactification scale --
whatever is the scale at which four-dimensional field theory
breaks down -- whether or not there is a unified embedding of the
Standard Model. Of course, for the coupling \corko\ of the axion
to electromagnetism, we must interpret $k$ to be the level of the
current algebra embedding of the electromagnetic $U(1)$.

The string scale $M_s=1/\ell_s$ can be evaluated from the above
formulas to be $M_s=(k\alpha_{G}/4\pi )^{1/2}M_P$.  With the usual
phenomenological estimate $\alpha_C\sim 1/25$, we get the usual
perturbative heterotic string scale $M_s\sim M_P\sqrt{k}/18.$

Axions arise from the $B$-field of the heterotic string.  The
components $B_{\mu\nu}$ with $\mu$ and $\nu$ tangent to Minkowski
spacetime (and constant on $Z$) can be dualized to make an axion.
Since the existence and properties of this axion do not depend
very much on the choice of $Z$, it has been traditionally called
the model-independent axion of the heterotic string.  Though the
name seems somewhat anachronistic in an age in which there are
many other string-based models, we will see that something rather
like this mode does exist in many other asymptotic limits of
string theory.  Zero modes of $B_{\mu\nu}$ with $\mu$ and $\nu$
tangent to the compact manifold $Z$ also have axion-like couplings
and are traditionally known as model-dependent axions.

\bigskip\noindent{\it Model-Independent Axion}

We consider first the model-independent axion.  The Bianchi
identity for the gauge-invariant field strength of $H$ is
\eqn\iglo{dH={1\over 16\pi^2}\left(\tr \,R\wedge R-\tr\, F\wedge
F\right).} (For example, the normalization can be extracted from
eqn. (12.1.40) of \polchinski, bearing in mind that our $H$ is
$\ell_s^2$ times the $H$-field used there.)  Now focus on the
modes that are constant on $Z$ with all indices tangent to
four-dimensional spacetime.  The four-dimensional component of the
$B$-field can be dualized by introducing a field $a$ that is a
Lagrange multiplier for the Bianchi identity, the coupling being
$\int a\left(dH+{1\over 16\pi^2}\left(\tr\,F\wedge F-\tr\,R\wedge
R\right)\right)$. Including also the $B$-field kinetic energy from
\hetef, the action is \eqn\jumbo{- {2\pi V_Z \over g_s^2 l_s^4}
\int d^4x\, {1\over 2}H\wedge \star H+\int a\left(dH+{1\over
16\pi^2}\left(\tr F\wedge F-\tr R\wedge R\right)\right) } Here $H$
is an independent field variable (which can be expressed in terms
of $B$ if one integrates first over $a$ to impose the Bianchi
identity). As $H$ has integer periods, $a$ should have period
$2\pi$.\foot{One can carry out the duality more precisely, in a
way that is valid on a general four-manifold, by adapting the
procedure used in \ref\rocver{M. Rocek and E. Verlinde, ``Duality,
Quotients, And Currents,'' Nucl. Phys. {\bf B373} (1992) 630,
hep-th/9110053.} for two-dimensional $T$-duality.} Instead, we
integrate out $H$ to get an effective action for $a$:
\eqn\toyo{S(a)={g_s^2 \ell_s^4\over 2\pi V_Z} \int d^4x\left(
-{1\over 2}\partial_\mu a
\partial^\mu a\right) + \int a {1\over 16\pi^2}\left(\tr F\wedge F-\tr R\wedge R\right).}
Since $\tr F\wedge F = 2k {\bf tr} F\wedge F$, we see that for
this particular axion, the integer $r$ characterizing the axionic
coupling is equal to the current algebra level $k$.  For future
reference, let us note that the effect of the dualization from $H$
to $a$ is that $F_a^2$ is simply the inverse of the coefficient of
$(1/2)H\wedge \star H$ in the four-dimensional kinetic energy of
$H$.

So we can read off the axion decay constant:
\eqn\olbo{F_a={g_s^2\ell_s^2\over \sqrt{2\pi
V_Z}}={k\alpha_{G}\over 2\pi}{M_P\over \sqrt 2}.} The axion
couplings are proportional, according to \umbo, to $F_a/k$, which
is \eqn\tumbo{{F_a\over k}= {\alpha_CM_P\over 2\pi\sqrt 2}.} If we
take $\alpha_C=1/25$, this gives $F_a/k\approx 1.1\times 10^{16}$
GeV, as in \thomas.  If the model is not unified or the spectrum
of particles contributing to the renormalization group running is
not the minimal one, then $\alpha_C$ may have a somewhat different
value, leading to a somewhat different value of $F_a/k$.

In order for the axionic mode whose coupling we have just
evaluated to solve the strong CP problem, low energy QCD
instantons must be the dominant mechanism that violates the
associated PQ symmetry. There are a few other candidates to worry
about.  Gauge instantons at the string scale have action
\eqn\ilko{{2\pi\over\alpha_{Y\negthinspace M}}={2\pi\over
k\alpha_C}\sim {157\over k},} by analogy with \instac.  For $k=1$,
this is somewhat below the value 200   that we need according to
\gogoo\ if there is no suppression due to low energy
supersymmetry. But it might be satisfactory if because of
supersymmetry a formula such as \ubu\ is more appropriate, or if
for some reason $\alpha_{Y\negthinspace M}$ is a bit smaller than
$1/25$.  For $k>1$, it would be much harder to suppress the
explicit PQ violation adequately.

Because of the $a\,\tr R\wedge R$ coupling, the PQ symmetry might
also be violated by gravitational instantons, if they are relevant
to string theory in asymptotically flat spacetime, but we do not
know how to estimate their effects. Finally, it might happen that
some other part of the $E_8\times E_8$ or $SO(32)$ gauge group of
the heterotic string becomes strong at an energy above the QCD
scale. Then instantons of that group might be the dominant source
of PQ symmetry violation.

If one of these sources of PQ violation gives an excessive mass to
the mode $a$, then we need another axion to solve the strong CP
problem.  Interestingly, there are other candidates, as we now
discuss.

\bigskip\noindent{\it Model-Dependent Axions}

Model-dependent heterotic string axions arise from zero modes of
the $B$-field on the compact manifold $Z$.  Let there be $n={\rm
dim}\,H^2(Z,\Bbb{R})$ such zero modes $\beta_1,\dots,\beta_n$. We
normalize them so that \eqn\toyo{\int_{C_j}\beta_i=\delta_{ij},}
where the $C_j$ are two-cycles representing a basis of
$H_2(Z,\Bbb{Z})$ modulo torsion.
 Then we make an ansatz \eqn\jin{B={1\over 2\pi}\sum_i
\beta_ib_i,} where $b_i$ are four-dimensional fields. The factor
of $1/2\pi$ is included so that the fields $b_i$ have periods
$2\pi$, as is conventional for axions.   Set
\eqn\tolfgo{\gamma_{ij}=\int_Z\beta_i\wedge *\beta_j.} By
dimensional reduction from the $B$-field kinetic energy in \hete,
the kinetic energy of the $b_i$ fields in four dimensions comes
out to be \eqn\undu{S_{kin}=-{1\over 2\pi g_s^2\ell_s^4}\int
d^4x{\gamma_{ij}\over 2}\partial_\mu b_i\partial^\mu b_j.}

Dimensional reduction of a {\it local} action in ten dimensions
leads to a four-dimensional effective action in which the modes
$b_i$ only have derivative couplings. This results from the
underlying gauge invariance $\delta B=d\Lambda$ in ten dimensions,
and  is why these modes have approximate PQ shift symmetries.

As first observed in \ewitten, these modes acquire axionic
couplings from the one-loop couplings that enter the Green-Schwarz
anomaly cancellation mechanism.  The relevant couplings are
\eqn\purgo{{-1\over 4(2\pi)^3 4!}\int B \left\{ -{\Tr F\wedge F
\tr R\wedge R\over 30}+{\Tr F^4\over 3}-{(\Tr F\wedge F)^2\over
900}\right\}.} (We have omitted some purely gravitational terms,
which lead to a coupling of the modes $b_i$ to $\tr R\wedge R$ in
four dimensions.) To proceed further, we consider the $E_8\times
E_8$ heterotic string, embedding the Standard Model in the first
$E_8$, and write $\tr_1$ and $\tr_2$ for traces in the first or
second $E_8$. The qualitative conclusions are not different for
the $SO(32)$ heterotic string or if the Standard Model embedding
in $E_8\times E_8$ is more complicated.  The couplings in four
dimensions of the axion modes to $\tr_1 F\wedge F$ come out to be
\eqn\jucoc{-{1\over 2\pi^2 4!}\sum_i\int_Z \beta_i\left\{-{\tr
R\wedge R\over 2}+ 2\tr_1 F\wedge F-\tr_2 F\wedge F\right\}\int_M
b_i {\tr_1 F\wedge F\over 16\pi^2}. } Using the Bianchi identity
\iglo, one can alternatively write these couplings as
\eqn\vucoc{-\sum_i\int_Z \beta_i \wedge{1\over 16\pi^2}\left(\tr_1
F\wedge F -{1\over 2}\tr R\wedge R\right)\int_M b_i{\tr_1 F\wedge
F\over 16\pi^2}.}

For a reasonably isotropic $Z$, a typical matrix element of the
metric $\gamma$ defined in \tolfgo\ is of order $V_Z^{1/3}$, and
hence a typical linear combination $b$ of the $b_i$ has $F_b\sim
V_Z^{1/3}/2\pi g_s^2\ell_s^4$.  If we use \fdc\ and \mormo\ to
eliminate $V_Z$ and $\ell_s$ in favor  of $M_P$ and $\alpha_C$, we
get \eqn\tongo{F_b={\alpha_C^{1/3}M_P\over 2\pi \sqrt
2k^{1/3}g_s^{2/3}}\grtsim {\alpha_C^{1/3}M_P\over 2\pi \sqrt 2
k^{1/3}}.} We assume in the last step that $g_s\lesssim 1$, for
validity of the computation.  For $k\sim 1$, $\alpha_C\sim 1/25$,
this gives a typical result $F_b\sim 10^{17}$ GeV.

To orient ourselves to how one might reduce $F_b$, let us consider
the case that $b_2(Z)=1$, so that there is a single harmonic
two-form $\beta$ and associated axion $b$. More specifically, we
take $Z=C\times Y$ where $C$ is a Riemann surface and $Y$ is a
four-manifold, which must be spin as the theory contains fermions.
(Of course, in supersymmetric compactifications, $Z$ is generally
not such a product and is considerably more complicated.) For the
mode $b$, the integer that appears in the $b \,\tr F\wedge F$
coupling to four-dimensional gauge fields is {\it not} the same as
the level $k$ of the current algebra embedding of the Standard
Model. Rather, from \vucoc, this integer is\foot{Despite the
factor of $1/2$ multiplying $\tr R \wedge R$, this expression is
an integer for $Y$ a spin manifold, because the signature of a
four-dimensional spin manifold is even and in fact divisible by
16.}
 \eqn\nurmy{k'={1\over 16\pi
^2}\int_Y\left(\tr_1 F\wedge F -{1\over 2}\tr R\wedge R\right).}

We write $V_C$ and $V_Y$ for the volumes of $C$ and $Y$. Then as
$\int_C\beta=1$, we have \eqn\wwh{\int_Z\beta\wedge\star
\beta=V_C^{-1}V_Y={V_Z\over V_C^2}.} Taking the $b$ kinetic energy
from \undu\ and $M_P$ from \fdc, we get
\eqn\urcd{F_b={\ell_s^2\over 2\pi V_C}{M_P\over \sqrt 2}.} We may
as well assume that $V_C \gtrsim \ell_s^2$ (otherwise we should
make a $T$-duality to a better description in which all dimensions
in $Z$ are sub-stringy in scale).  So we get an approximate upper
bound on $F_b$, \eqn\wurcd{F_b\lesssim {M_P\over 2\pi\sqrt
2}=2.7\times 10^{17}\,{\rm GeV},} which is larger than the value
for the model-independent axion and slightly larger than the
generic estimate \tongo.

To {\it lower} $F_b$, we can take $V_C$ large, but there is a
limit to how far we can go.  From \apl\ and $V_Z=V_CV_Y$, we have
\eqn\infine{{V_C\over \ell_s^2}={g_s^2\over
k\alpha_C}{\ell_s^4\over V_Y}.} For the model to have a
qualitatively correct description as a weakly coupled heterotic
string, we require $g_s\lesssim 1$.  To have a sensible
description in terms of compactification on a manifold $C\times
Y$, we require $V_Y\gtrsim \ell_s^4$.  If either of these
conditions fails, one would want to make a duality transformation
to a better description (in the second case, for instance, this
would be a $T$-duality transformation) and proceed from there.  So
the validity of our approximations requires \eqn\pinfine{{V_C\over
\ell_s^2}\lesssim {1\over k\alpha_C}.} Inserting this in \urcd, we
find that \eqn\nurcd{{F_b\over k'}\gtrsim {k\over k'} {\alpha_C
M_P\over 2\pi \sqrt 2}.}

For $k=k'$, this lower bound agrees precisely with the actual
value that we computed for the model-independent axion. To
minimize $F_b/k'$, we can take $k=1$ and try to take $k'$ large.
Let us see how far we can get if we assume a supersymmetric
compactification.  A Calabi-Yau three-fold $Z$ cannot quite be a
product $C\times Y$.  But it can be a fibration, with fibers
$Y={\rm K3}$, over $C=\Bbb{CP}^1$.  This is good enough to justify
the above formulas.  We let $N_1 $ and $N_2$ denote the instanton
numbers in the two factors of the $E_8\times E_8$ gauge bundle
over $Y$. Supersymmetry requires $N_1,N_2\geq 0$, and the Bianchi
identity for $H$ implies that $N_1+N_2=24$ (we use the fact that
$24=(1/16\pi^2)\int_{\rm K3}\tr R\wedge R$ is the Euler
characteristic of K3). Moreover, from  \nurmy, we have
$k'=N_1-12$, so $|k'|\leq 12$.

So we can get $F_b/k'$ as small as about $10^{15}\,{\rm GeV}$ with
this kind of model, by putting all instantons in the same $E_8$ to
get $k'=12$.  (In fact, this is the most traditional type of
heterotic string model.)  To get $F_b$ as small as this, we need
to make $V_C$ as large as possible.  There is another virtue in
doing so.  Because of the coupling $2\pi i\int_\Sigma B$ to the
string worldsheet $\Sigma$, a worldsheet instanton described by a
string wrapping around $C$ explicitly breaks the PQ symmetry of
this particular axion.  The action of such an instanton is, in our
conventions, $I=2\pi V_C/\ell_s^2$.  To suppress explicit PQ
violation, we should make $V_C$ large.  When the inequality in
\nurcd\ is saturated, we have $I=2\pi /k\alpha_C$, the same value
as for the model-independent axion. Again, for $k=1$ and with the
help of some suppression by low energy supersymmetry, the PQ
symmetry might be good enough to solve the strong CP problem.

We introduced this example as a convenient special case, but
actually, to get $F_b$ small by making a two-cycle $C$ large, a
fibration over $C$ is the natural case to consider.  Another
approach to making $F_b$ small is to leave $V_C/\ell_s^2$ of order
1 and try to  make $k'$ very large.  An obvious potential problem
with this comes from the formula \nurmy\ for $k'$.  If $V_Y\sim
\ell_s^4$ (since larger $V_Y$ makes $F_b$ larger) and $k'$ is
large, then $F$ or $R$ is large pointwise and the approximations
may be invalid. We consider some examples seeking to make $k'$
large in section 5. But first, we repeat our analysis up to this
point for the strongly coupled $E_8\times E_8$ heterotic string.

\newsec{Heterotic $M$-Theory}

To describe the strongly coupled $E_8\times E_8$ heterotic string,
we consider compactification of $M$-theory on $Z\times I$, where
$Z$ is a six-manifold and $I$ is an interval.  At each end of the
interval $I$ lives an $E_8$ gauge multiplet.  Our $M$-theory
normalizations have been defined in section 2.

If the interval $I$ is relatively short, the metric on $Z\times I$
is approximately a product metric.  In this case, we write $\pi
\rho$ for the length of $I$, $V_Z$ for the volume of $Z$, and
$V_7=V_Z\pi \rho$ for the volume of $Z\times I$.  (Lengths and
volumes, of course, are now computed using the $M$-theory metric.)
When $I$ becomes long, the metric on $Z\times I$ ceases to be a
product metric, and if the instanton numbers are not equal at the
two ends of $I$, then the volume of $Z$ is larger at one end than
the other \ref\wittensc{E. Witten, ``Strong Coupling Expansion Of
Calabi-Yau Compactification,'' Nucl. Phys. {\bf B471} (1996) 135,
hep-th/9602070.}.  We have the option of embedding the Standard
Model into either of the two ends. In the usual approach to
phenomenology with the strongly coupled heterotic string, the
Standard Model lives at the end where volume of $Z$ is larger. For
now, we assume that this is the case. (The other possibility is
treated in section 4.3.) We let $V_Z$ be the volume of $Z$ at the
end where the Standard Model lives. For the volume of $Z\times I$,
we write $V_7=V_Z \pi \rho \epsilon$, where $\epsilon$ takes
account of the shrinking of $Z$ and equals 1 if $\rho$ is small
and 1/2 if the volume of $Z$ varies linearly from $V_Z$ at one end
to zero at the other.

By dimensional reduction from \olygo\ and \olfog, we find
\eqn\humbo{\eqalign{M_P^2 & = {4\pi^2\rho V_Z \epsilon\over
\ell_{11}^9}\cr \alpha_{Y\negthinspace M} & = {\ell_{11}^6\over
V_Z}\cr}} The value written in \humbo\ for $\alpha_{Y\negthinspace
M}=g_{Y\negthinspace M}^2/4\pi=k\alpha_C$ is valid at either end
of $I$ as long as one uses the appropriate value of $V_Z$ at that
end; since we have assumed that $V_Z$ is the larger of the two
volumes, the formula has really been written for the end of $I$ at
which the gauge coupling is weaker. If the Standard Model lives at
the end with smaller volume of $Z$, the effects of warping can be
significant; we will discuss this case in a later subsection.

\subsec{Model-Independent Axion}

We will now compute the coupling parameter $F_a$ of the
model-independent axion. The model-independent axion comes from a
mode of $G=dC$ with one index tangent to $I$ and the other three
tangent to the four-dimensional spacetime $M$:
\eqn\lokko{G={dx^{11}\over \pi \rho}H.} With this normalization,
$H$, like $G$, has integer periods.  The kinetic term for the
four-dimensional field $H$ comes from Kaluza-Klein reduction of
\olygo: \eqn\hiena{-{2V_Z\epsilon\over
\rho\ell_{11}^3}\int_M{1\over 2}H\wedge \star H.}  As in the
discussion of \toyo, $H$ can be dualized to an axion $a$, with
$F_a$ equal to the inverse of the $H$-field kinetic energy in
\lokko: \eqn\dogo{F_a=\sqrt{\rho\ell_{11}^3\over 2
V_Z\epsilon}={k\alpha_C\over 2\pi\epsilon}{M_P\over \sqrt 2}.} For
small $\rho$, $\epsilon=1$ and the result for $F_a$ agrees with
the analogous formula for the weakly coupled heterotic string; for
large $\rho$, $F_a$ may be larger by a factor of 2 because of the
factor $1/\epsilon$.

This computation also gives us our first illustration of the fact
that  in many classes of model, it is hard to significantly reduce
$F_a$ by going to a model with large extra dimensions.  Heterotic
$M$-theory gives a model with a large fifth dimension if we simply
assume that the instanton numbers are equal at the two ends of
$I$, in which case $\rho$ can become very large, limited only by
experimental tests of Newton's law of gravity.  The above
computation gave a result for the coupling parameter $F_a/k$ that
can be expressed in terms of the observables $\alpha_C$ and $M_P$,
independent of $\rho$. (When the instanton  numbers are equal at
the two ends, the metric on $Z\times I$ is nearly a product and we
can set $\epsilon$ to 1.) Of course, the reason that this happened
is that the relevant axionic mode is a bulk mode, like the
graviton.

The discussion of effects other than low energy QCD instantons
that violate the shift symmetry of the model-independent axion
would be similar to what it was for weak coupling.  Obvious
candidates are string scale instantons with action
$2\pi/k\alpha_C$, gravitational instantons, and instantons in the
second $E_8$.

\subsec{Model-Dependent Axions}

The model-dependent axions appear in an ansatz for the $C$-field:
\eqn\goro{C=\sum_i \beta_i{b_i\over 2\pi}\wedge {dx^{11}\over \pi
\rho}.}  The $\beta_i$ are defined as in \toyo, and as before the
$b_i$ are periodic scalars with period $2\pi$ and approximate
shift symmetries.

The couplings of these modes to $\tr\, F\wedge F$ and $\tr R\wedge
R$ in four dimensions are the same as they were for weak string
coupling. In fact, those couplings, being integers, are
independent of $g_s$.  From the standpoint of heterotic
$M$-theory, these couplings are evaluated by reducing to four
dimensions the term $(2\pi/6)\int C\wedge G\wedge G$ in the
$M$-theory effective action, and using the boundary condition
\ref\hw{P. Horava and E. Witten, ``Eleven-Dimensional Supergravity
On A Manifold With Boundary,'' Nucl. Phys. {\bf B475} (1996) 94,
hep-th/9603142.} that $G|_{x^{11}=0}=(\tr_1 F\wedge F -{1\over
2}\tr R\wedge R)/16\pi^2$, $G|_{x^{11}=\pi\rho}=-(\tr_2 F\wedge F
-{1\over 2}\tr R\wedge R)/16\pi^2$).  Upon making the reduction,
one finds, just as for the weakly coupled heterotic string, that
the relevant axionic couplings are \eqn\miglo{\eqalign{&\sum_i
\int_Z\beta_i\wedge {1\over 16\pi^2}\left(\tr_1 F\wedge F -{1\over
2}\tr R\wedge R\right)\int_Z b_i \tr F\wedge F \cr &= -\sum_i
\int_Z\beta_i\wedge {1\over 16\pi^2}\left(\tr_2 F\wedge F -{1\over
2}\tr R\wedge R\right)\int_Z b_i \tr F\wedge F.\cr}} The two
expressions are equal because of the Bianchi identity, which
implies that $[\tr_1 F\wedge F] +[\tr_2 F\wedge F] - [\tr R\wedge
R]=0$ (we write $[\alpha]$ for the cohomology class of a closed
differential form $\alpha$).

The Bianchi identity tells us that $[\tr_1 F\wedge F - \half \tr
R\wedge R]=0$ if and only if $[\tr_1 F\wedge F]=[\tr_2 F\wedge F]$
, that is, if and only if the two $E_8$ bundles over $Z$ are
topologically equivalent. Precisely when this is so, it is
possible in heterotic $M$-theory for the eleventh dimension to be
extremely long, giving a simple example of a model with a large
extra dimension.  But in this model, the phrase ``model-dependent
axions''  is a misnomer, as the modes in question do not have
axionic couplings.  When $[\tr_1 F\wedge F]\not= [\tr_2 F\wedge
F]$, the $b_i$ do have axionic couplings, and the length of the
eleventh dimension has an upper bound, which was found in
\wittensc\ and which we will discuss later.

By dimensional reduction, ignoring the variation of the geometry
with $x^{11}$, the kinetic energy of the fields $b_i$ is
\eqn\juju{{\epsilon\over 2\pi^2\ell_{11}^3\rho}\int_M{1\over
2}\gamma_{ij}^M\partial b_i\wedge \star
\partial b_j,}
where now \eqn\hico{\gamma_{ij}^M=\int_Z\beta_i\wedge \star
\beta_j.} The integral is evaluated at the end where the volume of
$Z$ is greater.

If $Z$ is fairly isotropic, matrix elements of $\gamma_M$ are of
order $V_Z^{1/3}$ and so the kinetic energy for a generic linear
combination of the $b_i$ is proportional to $F_b^2\sim \epsilon
V_Z^{1/3}/2\pi^2\ell_{11}^3\rho$.  As will be clear from our
evaluations below, this leads to $F_b$ being above the GUT scale
(except in the case that the $b_i$ lack the couplings of axions).
As in the weakly coupled case, we can try to make $F_b$ small for
one mode by taking a two-cycle to be large. We can also try to
make $F_b$ small by making $\rho$ large. We will consider first
the case of a large two-cycle, and show that it is not really
independent as it forces us to try to make $\rho$ large.

So we again consider the case that $Z$ is fibered over a Riemann
surface $C$, with fiber $Y$. The volumes then factorize:
$V_Z=V_CY_Y$. Moreover, for the axionic mode that is a pullback
from $C$, the relevant component of the metric is
\eqn\gbo{\int_Z\beta\wedge\star \beta ={V_Z\over V_C^2}.} For this
axion, we get then \eqn\nbo{F_b={M_P\ell_{11}^3\over 2\pi^2\sqrt 2
V_C\rho}.} We can try to make $F_b$ small by making $V_C$ large,
but as in the weakly coupled case, there is a limit to how far one
can go in that direction.  Since
$k\alpha_C=\ell_{11}^6/V_Z=\ell_{11}^6/V_YV_C$, we have
$1/V_C=k\alpha_CV_Y/\ell_{11}^6$.  The $M$-theory description only
makes sense if $V_Y\gtrsim \ell_{11}^4$, so $1/V_C\gtrsim
k\alpha_C/ \ell_{11}^2$. Hence \eqn\jobbo{F_b\gtrsim{
{k\alpha_C}\over 2\pi}{M_P\over \sqrt 2}{\ell_{11}\over \pi
\rho}.} So we cannot make much progress in reducing $F_b$ by going
to large $V_C$, unless $\rho$ is also large.

Let us then discuss what happens upon taking $\rho$ large. If
$[\tr_1 F\wedge F]=[\tr_2 F\wedge F]$, then in heterotic
$M$-theory,  $\ell_{11}/\pi\rho$ can be many orders of magnitude
less than 1, and $F_b$ can be in the range allowed by the usual
cosmological arguments, $10^9\,{\rm GeV}\lesssim F_b\lesssim
10^{12}$ GeV, or much lower. However, precisely in this case, as
we have already discussed,  the modes $b_i$ lacks axionic coupling
to $\tr F\wedge F$.

So if we want to solve the strong CP problem using some of these
modes as axions, we have to assume that $[\tr_1 F\wedge F]\not=
[\tr_2 F\wedge F]$.   That being so, there is an upper bound on
$\pi\rho/\ell_{11}$, which \wittensc\ cannot be larger than
roughly $\alpha_C^{-2/3}$ times a number of order one.  (Moreover,
reasonable GUT phenomenology can arise \refs{\wittensc,\bd} when
$\pi\rho$ is near the upper bound.) Because the axion kinetic
energies \juju\ or \nbo\ are proportional to $1/\rho$, this
enables us to suppress the $F_{b_i}$  by about an order of
magnitude relative to the familiar value $\alpha_CM_P/2\pi\sqrt
2\sim 1.1\times 10^{16}$ GeV.

The upper bound on $\rho$ comes from the way the metric on $Z$
varies as a function of the eleventh dimension. For example, if the
volume of $Z$ is going to zero at one endpoint of $I$ (which is not
necessarily so as the extended Kahler cone has other boundaries),
then the upper bound on the possible value of $\pi\rho$ was very
roughly, within perhaps a factor of 2, estimated in \wittensc\ to be
\eqn\ilfug{\pi\rho_{max}={V_Z\over
\ell_{11}^3\left\vert\int_Z\omega\wedge {\tr F\wedge F -{1\over
2}\tr R\wedge R\over 16\pi^2}\right\vert},} where $\omega$ is the
Kahler class of $Z$.  Under an overall scaling of $Z$, the integral
in the denominator scales as $V_Z^{1/3}$, so
$\pi\rho_{max}/\ell_{11}$ scales as
$V_Z^{2/3}/\ell_{11}^4\sim\alpha_C^{-2/3}$.  In the example that $Z$
is a K3 fibration over a large two-cycle $C$, if we assume that the
dominant contribution to the integral in the denominator is
$\int_Y(\tr_1 F\wedge F-{1\over 2}\tr R\wedge R)/16\pi^2
\cdot\int_C\omega$, then we can estimate the integral as
$V_C|N_1-12|=V_C|k'|$, so \eqn\jugy{\pi\rho_{max}V_C\lesssim
V_Z/\ell_{11}^3|k'|.}  When this is inserted in \nbo, we get
$F_b/|k'|\gtrsim {k\alpha_C M_P/2\pi\sqrt 2}\sim k\cdot 1.1\times
10^{16}$ GeV, a familiar value from the heterotic string. What has
happened is simply that since $\rho$ and $V_C$ vary reciprocally, we
cannot make them both large.

\bigskip\noindent{\it Explicit Violation Of The PQ Symmetries}

The PQ symmetry of these modes is violated explicitly by membrane
instantons wrapped on $D\times I$, where $D$ is a two-cycle in
$Z$.  The volume of such a membrane is $V_D
=\pi\rho\tilde\epsilon$ (where $\tilde\epsilon$ is a factor
similar to $\epsilon$ and accounts for how the area of $D$ varies
in the eleventh dimension). As the M2-brane tension is
$2\pi/\ell_{11}^3$, the action is $S=2\pi^2 \rho V_D/\ell_{11}^3$.
For a fairly isotropic $Z$,  $V_D/\ell_{11}^2\sim
(V_Z/\ell_{11}^6)^{1/3}\sim \alpha_C^{-1/3}$, and
$\pi\rho_{max}/\ell_{11}\sim\alpha_C^{-2/3}$, so $S\sim
2\pi/\alpha_C$.  Hence, explicit violation of the PQ symmetries
might be small enough to solve the strong CP problem.

For the case of a fibration $Z\to C$ with a large two-cycle $C$,
our inequality \jugy\ together with \humbo\ leads to
$S=2\pi\tilde\epsilon/\alpha_C|k'|$.  So we might need $|k'|=1$ to
have a sufficiently good PQ symmetry for this mode.

\subsec{Warped Compactification}

In heterotic $M$-theory, the metric on $M\times Z\times I$ is
actually a warped product, not an ordinary product, but we have
not taken this into account so far.
 Here, we study the question of whether warping
could play an important effect in heterotic $M$-theory. As we will
see, the warping does not change the qualitative conclusions in
the usual approach to phenomenology in which the Standard Model is
enbedded at the end of $I$ at which the volume is larger. However,
it can be quite important in the opposite case.

\nref\curioa{G. Curio and A. Krause,``Four-Flux and Warped
Heterotic $M$-theory Compactifications," Nucl. Phys. {\bf B602}
(2001) 172, hep-th/0012152.} \nref\curiob{G. Curio and A. Krause,
``Enlarging the Parameter Space of Heterotic $M$-theory Flux
Compactifications to Phenomenological Viability," Nucl. Phys. {\bf
B693} (2004) 195, hep-th/0308202.} In supersymmetric
compactification of heterotic $M$-theory, the warped metric takes
the simple form \refs{\wittensc,\curioa,\curiob}
\eqn\warpm{ds^2=e^{-f(x_{11})} \eta_{\mu\nu} dx^\mu dx^\nu
+e^{f(x_{11})}\l( g_{mn} dy^m dy^n+ dx^{11} dx^{11}\r).} Here
$y^m$ are  local coordinates on $Z$ and $x^{11}$ is parameterizes
the heterotic $M$-theory interval.  The warp factor is
\eqn\warpf{e^{f(x_{11})}=(1+ x^{11}Q)^{2/3}.} (Without assuming
low energy supersymmetry, more general fluxes are possible, but
the supersymmetric case seems general enough to illustrate our
point.)  The parameter $Q$, which we will loosely call ``instanton
number,'' is given by \eqn\quv{Q={\ell_{11}^3\over 2 V_Z}\int_Z
\omega\wedge {1\over 16\pi^2}\l(\tr_1 F\wedge F-\half \tr R\wedge
R\r),} where $\omega$ is the Kahler form of $Z$. Note that, if the
instanton number at $x^{11}=0$ is larger than at $x^{11}=\pi\rho$,
then the integral \quv\ is negative. This follows from the
supersymmetry relation $\omega_{ij} F^{ij}=0.$ The integral is
roughly $V_Z^{1/3}=\ell_{11}^2(k\alpha_{C})^{-1/3}$, so we express
it in terms of a dimensionless number $q$ of order one
\eqn\parq{Q={(k\alpha_{C})^{2/3}\over 2\ell_{11}}q.} If the
instanton number at the $x^{11}=0$ boundary is greater than the
instanton number at the other boundary, $Q$ is negative and $Z$
shrinks along $x^{11}.$ In this case, there is a critical
coordinate distance \eqn\crico{\pi\rho_{max}={1\over |Q|}} at
which the warp factor \warpf\ becomes zero and our supergravity
approximation breaks down. Hence, in the following we assume that
$\rho<\rho_{max}$. In the opposite situation with $Q$ positive,
$Z$ expands along the interval so the length of the interval could
be in principle arbitrarily large.

Dimensional reduction of the gravity action \olygo\ to four
dimensions gives \eqn\gravir{ S={2\pi V_Z\over \ell_{11}^9}
\int_0^{\pi\rho}dx^{11} e^{5/2 f(x^{11})}\int d^4x \sqrt{-g}
R,
} whence the four-dimensional Planck mass is \eqn\mpli{M_P^2={3\pi
V_Z\over 2\ell_{11}^9Q}\l[(1+\pi\rho Q)^{8/3}-1\r].}

The gauge fields of the two $E_8$'s live on the two ends of the
interval. Reducing the Yang-Mills action \olfog\
\eqn\ymrem{S_{Y\negthinspace M}=-{1\over
4(2\pi)\ell_{11}^6}\int_{M^{10}} \tr F\wedge \star F} to four
dimensions gives the gauge coupling as a function of the boundary
location \eqn\ymdep{{1\over\alpha_{Y\negthinspace
M}(x^{11})}={V_0\over\ell_{11}^6}(1+x^{11} Q)^2.} In the linear
approximation this agrees with the result of \wittensc\
\eqn\volv{{d\over d x^{11}}\l({1\over \alpha_{Y\negthinspace
M}}\r)=\ell_{11}^{-3}\int_Z\omega\wedge \l(\tr_1 F\wedge F-\half \tr
R\wedge R\r).}

\medskip\noindent{\it Standard Approach to Phenomenology}

Now let us re-examine the case that the Standard Model is embedded
in the larger of the two boundaries. This will give results
qualitatively similar to those we found above, but somewhat more
precise. With the Standard Model embedded at the larger end, the
volume is decreasing away from the Standard Model boundary. At a
coordinate distance $\pi\rho_{crit}=1/|Q| $ the volume becomes zero
in our approximation. This gives an upper bound on $M_P$ \wittensc.\
Substituting $1/|Q|$ for $\pi\rho$ into the expression for the
Planck mass we find \eqn\mpmax{M_{P,max}^2={1\over
\ell_{11}^2}{3\pi\over q(k\alpha_C)^{5/3} },} which is a factor of
$3/2$ larger \curioa\ than the limit on $M_P^2$ obtained in the
linear approximation \refs{\wittensc}.

The warping of the metric \warpm\ affects the internal wavefunction
of the axion as well. The warping is different for the
model-independent and for the model-dependent axions. For
model-independent axion we modify the ansatz \lokko\ by a nontrivial
warp factor \eqn\indem{C=\alpha e^{g(x^{11})} B(x^\mu)\wedge
dx^{11},} where $\alpha$ is a normalization constant and $g(x^{11})$
captures the dependence of $C$ on $x^{11}.$ To find $g$ we require
that the three-form equation of motion $d\star d C=0$ in eleven
dimensions reduces to the four-dimensional equation of motion $
d_4\star_4 d_4 B=0$ for the two-form field $B$. We substitute
\indem\ into the equation of motion $d\star d C=0$ which, after
taking the first exterior differential and the Hodge dual becomes
\eqn\hame{ d\l(e^{g+{7\over2}f}d\,{\rm vol}_Z\wedge(\star_4 d_4
B)\r)=0,} where $d\,{\rm vol}_Z$ is the volume form on $Z$.
Expanding the second exterior differential into its four-dimensional
and seven-dimensional parts $d=d_4+d_7$, \hame\ becomes
\eqn\rame{d_4\star_4 d_4 B+d_7(g+{7\over2}f)\star_4 d_4 B=0.} In
writing this we suppressed an overall factor of $e^{g+7f/2} d\,{\rm
vol}_Z.$ The equation\rame\ reduces to the four-dimensional equation
of motion for the massless two-form $B$ if $g=-7f/2+c$. Here $c$ is
a constant that we absorb into the normalization constant $\alpha$
of the $C$-field. Hence the ansatz for the model-independent axion
is \eqn\cform{C=\alpha(1+x^{11}Q)^{-7/3}B\wedge dx^{11}.} We fix the
normalization constant $\alpha$ by requiring $G=dC$ to have integer
periods. We assume that $B$ is normalized so that $H=dB$ has integer
periods, so that the axion has period $2\pi.$ With this
normalization, $G$ will have integer periods if $\alpha
(1+x^{11}Q)^{-7/3} dx^{11}$ integrates to one on the $M$-theory
interval $I$. This condition fixes $\alpha$ to be
\eqn\alpd{\alpha^{-1}
={3\over4Q}\l[1-(1+\pi\rho Q)^{-4/3}\r].} Dimensional reduction of
the $G$-field kinetic energy \olygo\ gives the kinetic term for
the four-dimensional $B$-field \eqn\kinap{{2\pi \alpha V_Z\over
\ell_{11}^3}\int\left(-\half H\wedge \star H\right).} In four
dimensions, $B$ is dual to an axion with axion decay constant
equal to the inverse of the coefficient of the $H$-field
kinetic energy \eqn\apin{F_b^2
={1\over \ell_{11}^2}{3(k\alpha_C)^{1/3}\over 4\pi q}\l[1-(1+\pi\rho
Q)^{-4/3}\r].} From \apin\ we read off the dependence of $F_b$ on
the length of the $M$-theory interval. We assume that $Q$ is
negative, so that $Z$ gets smaller away from the Standard Model
boundary. When the interval is short, this agrees with the formula
\dogo\ that does not take into account warping. As the interval gets
longer, both $F_b$ and $M_P$ grow compared to $\ell_{11}^{-1}.$ At
$\pi\rho_{max}=1/|Q|,$ $M_P$ reaches its maximum value \mpmax\ while
$F_b$ becomes formally arbitrarily large. Intuitively, this is
because the wavefunction of the model-independent axion is
concentrated towards the $x^{11}=\pi\rho$ end of the interval, as is
clear from \cform,\ so the axion couples weakly to the Standard
Model fields which are supported on the other end of $I$.
\lref\VafaXN{
  C.~Vafa,
  ``Evidence for F-Theory,''
  Nucl.\ Phys.\ B {\bf 469}, 403 (1996)
  [arXiv:hep-th/9602022].
}
\lref\MorrisonNA{
  D.~R.~Morrison and C.~Vafa,
  ``Compactifications of F-Theory on Calabi--Yau Threefolds -- I,''
  Nucl.\ Phys.\ B {\bf 473}, 74 (1996)
  [arXiv:hep-th/9602114].
}
\lref\MorrisonPP{
  D.~R.~Morrison and C.~Vafa,
  ``Compactifications of F-Theory on Calabi--Yau Threefolds -- II,''
  Nucl.\ Phys.\ B {\bf 476}, 437 (1996)
  [arXiv:hep-th/9603161].
} However, before $F_b$ becomes super-Planckian, the size of $Z$ at
the $x^{11}=\pi\rho$ end of the interval becomes sub-Planckian and
our low energy supergravity description breaks down. This regime
should be studied in a well behaved dual description, such as
F-theory \refs{\VafaXN,\MorrisonNA,\MorrisonPP}.  We estimate the
maximum value of $\pi\rho$ for which we can trust our formula for
$F_b$ by requiring that the volume of $Z$ is at least one in
eleven-dimensional Planck units. From the warped metric \warpm\ we
read off the dependence of the volume of $Z$ on $x^{11}$ to be
$V=\ell_{11}^6(1+x^{11} Q)^2/(k\alpha_C)$. Requiring this to be at
least $\ell_{11}^6$ bounds the length of $I$: $(1+\pi\rho_{max}
Q)\lesssim(k \alpha_C)^{1/2}.$ Substituting this into \apin\ we
estimate an upper bound on the decay constant of the
model-independent axion \eqn\sapin{F_{b}\lesssim{1\over
\ell_{11}}\sqrt{3\over 4\pi q (k\alpha_C)^{1/3}}={k\alpha_C\over
2\pi\sqrt{q}}M_P.} Up to a factor of order one, this agrees with the
formula \dogo\ that does not take into account warping. Hence, the
warping does not modify the scale of the axion coupling parameter
significantly, when the Standard Model is embedded into the larger
of the two boundaries of $I.$

Let us now find the warp correction to the axion decay constant of
the model-dependent axions. We start by modifying the ansatz \goro\
for the three-form field with a warp factor
\eqn\zerom{C=\alpha\sum_i e^{g(x^{11})}{b_i\over 2\pi}\beta_i\wedge
 dx^{11}.} Here $\alpha$ is an normalization constant
that we will determine later. To find the warp factor $g(x^{11})$,
we substitute \zerom\ into the equation of motion $d\star d C=0$
which, after taking the first exterior derivative and the Hodge
star operator becomes \eqn\intere{\sum_i d\l(\star_4 d_4
b_i\wedge\star_6 \beta_i\, e^{g-\half f}\r)=0.} The symbols
$\star_4,\star_6$ denote the four and six-dimensional Hodge
operators. We simplified the equation using the fact that
$\beta_i$ are closed. \intere\ reduces to a set of
four-dimensional equations of motion for massless scalars $b_i$ if
$g=f/2$ plus a constant that we can absorb into the normalization
constant $\alpha.$ So the model-dependent axions come from the
ansatz \eqn\depo{C=\alpha\sum_i (1+x^{11}Q)^{1/3}{b_i\over 2\pi}
\beta_i\wedge dx^{11}.} The normalization constant $\alpha$ is
fixed by requiring that each three-form $\alpha (1+x^{11} Q)^{1/3}
\beta_i\wedge dx^{11}$ has unit flux through $C_i\times I.$ Since
$\int_{C_i}\beta_i=1,$ this gives
\eqn\appn{\alpha^{-1}=
{3\over 4Q}[(1+\pi\rho Q)^{4/3}-1].} Dimensional reduction of the
$C$-field kinetic energy \olygo\ leads to the kinetic terms of the
axions $b_i$ \eqn\kinoma{{\alpha\over 2\pi\ell_{11}^3}\int_M-\half
\gamma_{ij}^M\partial b_i\wedge \star\partial b_j} with
$\gamma_{ij}^M$ given by \hico.\

In the standard approach to phenomenology, the volume of $Z$
decreases away from the end of the interval with Standard Model
fields. We saw already in the linear approximation \nbo\ that
$F_b$ decreases as we increase the length of the interval $I$.
Hence, to find the smallest possible $F_b$  we take $\rho$ large,
$\pi\rho_{max}=|Q|^{-1}.$ For an axion coming from a generic cycle
$C$, we estimate the integral $\int_Z \beta\wedge\star \beta\sim
V_Z^{1/3}=\ell_{11}^2(k\alpha_C)^{-1/3},$ so the axion decay
constant in terms of the Planck mass \mpmax\ is
\eqn\modls{F_b\gtrsim{qk\alpha_C\over3\pi} M_P,} which is close to
the familiar value $1.1\times 10^{16}\GeV.$ As we discussed
before, we can try to make $F_b$ small by taking size of the cycle
$C$ large. So we take $Z$ to be a fibration over a large Riemann
surface $C$ with a fiber $Y.$ A calculation identical to the one
performed in section 4.2 shows that taking $C$ large does not help
in lowering $F_b$ because $V_C$ and $\rho_{max}$ vary
reciprocally, so we cannot make them both large.

\medskip\noindent{\it Non-Standard Approach to Phenomenology}

The warping does not modify the scale of the axion coupling
parameter significantly, when the Standard Model is embedded into
the larger of the two boundaries of $I.$ On the other hand we
expect the effects of the warping to be pronounced if the size of
$Z$ is increasing away from the Standard Model boundary. In this
case, the length of the interval can be arbitrarily large. As we
increase it, the volume of the seven-dimensional compactification
manifold gets larger so the four-dimensional Planck scale grows
compared to the $M$-theory scale $\ell_{11}^{-1}$. This is clear
from the formula \mpli\ for $M_P$  with $Q$ taken to be positive.

The wavefunction of the model-independent axion \cform\ is localized
near the SM end of the interval, so the axion decay constant stays
close to the $M$-theory scale $\ell_{11}^{-1}$. Indeed, in the limit
of large interval $\pi\rho\gg Q^{-1},$ the axion decay constant
\apin\ approaches a constant multiple of $\ell_{11}^{-1}$
\eqn\alip{F_b^2={1\over \ell_{11}^2}{3(k\alpha_C)^{1/3}\over4\pi
q},} while the Planck mass \mpli\ grows with $\pi\rho$ as
\eqn\mplag{M_P^2={1\over \ell_{11}^2}{3\pi(\pi\rho Q)^{8/3}\over
q(k\alpha_C)^{5/3}}.} Hence, increasing $\rho$ lowers the axion
decay constant compared to $M_P$ \eqn\flowp{F_b={k\alpha_C\over
2\pi} {M_P\over (\pi\rho Q)^{4/3}}.} Solving \alip\ and \mplag\ for
$M_{11}$ and $\pi\rho,$ we get
\eqn\moso{\eqalign{M_{11}&=F_b\sqrt{4\pi q\over
3(k\alpha_C)^{1/3}},\cr \pi\rho &=\ell_{11}\l({M_P\over
F_b}\r)^{3/4}{2^{1/4}(k\alpha_C)^{1/12}\over\pi^{3/4}q}.}} For
phenomenologically preferred axion decay constant $10^9\GeV<
F_b<10^{12}\GeV,$ the allowed range of the $M$-theory parameters is
\eqn\repi{\eqalign{3.5\times 10^9\GeV\lesssim & M_{11}\lesssim
3.5\times10^{12}\GeV,\cr 4.2\times10^6\ell_{11}\gtrsim &\rho \gtrsim
2.3\times 10^4\ell_{11}.}}

Let us now consider a generic model-dependent axion with
$\int_Z\beta\wedge \star\beta\sim V_Z^{1/3}.$ We get the axion
decay constant by substituting for the normalization constant
$\alpha={4Q\over3}(\pi\rho Q)^{-4/3}$ from \appn\ into \kinoma\
\eqn\axisi{F_b=M_P{4\over 3\pi
q(k\alpha_C)^{1/3}}\l({\ell_{11}\over \pi\rho}\r)^2.} To simplify
\axisi\ we used that $Q= q(k\alpha_C)^{2/3}/2\ell_{11}$,
$V_Z=\ell_{11}^6/(k\alpha_C)$ and the expression \mplag\ for
$M_P$. It is clear from the formula for $F_b$ that as the length
of $I$ gets larger, the axion decay scale is parametrically
lowered compared to $M_P.$ We can estimate how much is it
necessary to elongate the interval to bring $F_b$ down to the
cosmologically favored region. From \axisi\ and \mplag\ we express
$\pi\rho$ and $M_{11}=\ell_{11}^{-1}$ as
\eqn\arpa{\eqalign{M_{11}&=F_b\l({M_P\over F_b}\r)^{1/3}\l({3\pi
k\alpha_C\over q}\r)^{1/6} \cr \pi\rho &=\ell_{11}\l({M_P\over
F_b}\r)^{1/2} {2\over \sqrt{3\pi q (k\alpha_C)^{1/3}}}.}} If we
take $k,q=1,$ then the axion decay constant takes the
phenomenologically preferred values $10^9\GeV<F_b< 10^{12}\GeV$
for the $M$-theory scale and the length of the interval lying in
the range \eqn\pira{\eqalign{ 1.1\times 10^{12}\GeV\lesssim
&M_{11}\lesssim 1.1\times 10^{14}\GeV \cr 5.4\times 10^4
\ell_{11}\gtrsim &\pi\rho\gtrsim 1.7\times 10^3\ell_{11}.}}

\medskip\noindent {\it Generalization to Other Warped
Compactifications}

\nref\porrati{N. Arkani-Hamed, M. Porrati and L. Randall,
``Holography and Phenomenology,'' JHEP 0108:017 (2001)
hep-th/0012148.}

In the context of heterotic $M$-theory, we have seen how significant
warping can lower the axion decay scale. A similar effect can occur
in other warped compactifications, for example in compactifications
of the type II superstring. If the axion is supported near a region
with significant warping, its decay constant will be lowered
compared to the Planck scale just like in the case of heterotic
$M$-theory with non-standard approach to phenomenology.

\newsec{Anomalous $U(1)$ Symmetries in String
Theory}

\nref\atick{J. J. Atick, L. J. Dixon and A. Sen, ``String
Calculation of Fayet-Iliopoulos $D$ Terms in Arbitrary
Supersymmetric Compactifications,'' Nucl. Phys. {\bf B292} (1987)
109.} \nref\fiterms{M. Dine, N. Seiberg and E. Witten,
``Fayet-Iliopoulos Terms in String Theory,'' Nucl. Phys. {\bf
B289} (1987) 589.} \nref\fdterms{M. Dine, I. Ichinose and N.
Seiberg,``$F$ Terms and $D$ Terms in String Theory,'' Nucl. Phys.
{\bf B293} (1987) 253.} \nref\russell{J. March-Russell,``The
Fayet-Iliopoulos Term in Type I String Theory and $M$-theory,''
Phys. Lett. {\bf B437} (1998) 318, hep-ph/9806426.}
\nref\dinehamed{N. Arkani-Hamed, M. Dine and S. P.
Martin,``Dynamical Supersymmetry Breaking in Models with a
Green-Schwarz Mechanism,'' Phys. Lett. {\bf B431} 329,
hep-ph/9803432.}

To discuss the role of anomalous $U(1)$ symmetries for axion
physics, we begin by considering compactifications of the heterotic
string on a smooth six-manifold. In many such compactifications, the
low energy gauge group, understood as the subgroup of $E_8\times
E_8$ or $SO(32)$ that commutes with the gauge field expectation
value on the compact manifold,  contains an anomalous abelian gauge
symmetry $U(1)_B$ (or several such symmetries). For example, this is
very common in supersymmetric $(0,2)$ models in which the gauge
fields on the compact manifold have $U(1)$ factors. The anomaly
appears in the charges of the massless fermions. It is canceled by a
Green-Schwarz mechanism involving one of the axion multiplets.  For
brevity, we will consider the case that the axion in question is the
model-independent one. The relevant fields participating in the
four-dimensional Green-Schwarz mechanism are in that case the vector
field $V_B$ of the anomalous $U(1)_B$ and the axion-dilaton field
$S=1/g_{B}^2+ia/8\pi^2,$ where $g_B$ is the four-dimensional gauge
coupling of $U(1)_B$. The Kahler potential for these fields is
\refs{\fiterms,\fdterms,\atick}
\eqn\kapot{K=-M_P^2\ln(S+\overline{S}-c V_X),} where $c={\tr B/6}$
is a multiple of the chiral trace of the generator of $U(1)_B$ over
the massless fermions. The chiral trace counts right-handed fields
with an extra minus sign compared to left-handed ones. Under a gauge
variation $V_B\rightarrow V_B+i(\Lambda_B-\overline{\Lambda}_B),$
the axion-dilaton superfield transforms as $S\rightarrow
S+ic\,\Lambda_B.$ Hence the axion has shift symmetry
\eqn\shifts{a\rightarrow a+c\,\theta_B,} where $\theta_B=\Re
\Lambda_B|_{\theta=\overline\theta=0}.$ This makes possible the
anomaly cancellation. The axion has anomalous couplings $a\, \tr
F_i\wedge F_i$ to nonabelian gauge fields; the variation of these
couplings under \shifts\ exactly cancels the $U(1)_B$ anomaly.

We can now study the four-dimensional effective theory of the
axion and the $U(1)_B$ gauge fields. The effective action contains
the terms \eqn\kipot{ -\half F_a^2\l(\partial_\mu a+B_\mu{\tr
B\over 12}\r)^2+ \zeta^2 D_B+a {1\over 16\pi^2}\tr\,( F\wedge
F)_{\QCD},} where the first two terms come form expanding the
Kalher term \kapot\ and $\tr\,(F\wedge F)_{\QCD}$ is a multiple of
the QCD instanton density. $F_a=k\alpha_C M_P/(2\pi\sqrt{2})$ is
the familiar axion decay constant of the model-independent axion
and $\zeta$ is the Fayet-Iliopoulos term
\eqn\fate{\zeta^2=k\alpha_C M_P^2{\tr B\over 48\pi}.} {}From the
effective action \kipot, we see that the gauge boson eats the
axion and acquires a mass via the Higgs mechanism:
\eqn\gamas{M_B={k\alpha_C\over \ell_s}{\tr B\over 6\sqrt{2}}.}
Below the scale $M_B$, the anomalous gauge boson decouples,
leaving behind, in sigma model and spacetime perturbation theory,
an anomalous global $U(1)_B$ symmetry \ewitten.\  This symmetry is
broken explicitly by instantons.

It has been recognized in \BarrHK\ that the surviving global
symmetry can be used to solve the strong CP-problem with $F_a$
well below the string scale. In reducing to four dimensions, the
four-dimensional massless spectrum frequently contains scalars,
i.e. from the gauge bundle moduli, that are charged under the
anomalous symmetry and are neutral under other gauge groups. For
illustration, we assume that there is one such scalar field $\phi$
with charge $q$ under $U(1)_B.$ If $\phi$ acquires a VEV
$\langle\phi\rangle$, this spontaneously breaks the global
$U(1)_B$. We will assume that this VEV results from some dynamics
at an energy scale below the string scale, so that
$|\langle\phi\rangle|<M_B$. The $U(1)_B$ gets realized nonlinearly
by shifting the phase of $\phi=|\phi|e^{ib}$
\eqn\nore{b\rightarrow b+q\theta_B.} The kinetic energy of $b$
follows from the $\phi$ kinetic term $-|D_\mu \phi|^2$
\eqn\bkine{-|\phi|^2(\partial_\mu b-qA_\mu).} Hence, $b$ is a PQ
axion with decay constant $F_b=\sqrt{2}|\phi|.$ Its coupling to
the QCD instanton density is determined by the underlying $U(1)_B$
anomaly to be \eqn\bcom{b{c\over 16\pi^2q}\tr \,(F\wedge
F)_{\QCD}} with axionic coupling $k=c/q.$ The couplings of this
axion to matter are determined by the ratio $F_b/k$.  This is
\eqn\acip{{F_b\over k}= {12\sqrt{2}q \over \Tr B}|\phi|,} which is
roughly $|\phi.|$ Hence, if $\phi$ can be stabilized with
expectation value much less than the string scale, the axion might
have decay constant in the favored range
$10^9\GeV<F_b<10^{12}\GeV.$

In supersymmetric compactifications of the heterotic string, it
seems difficult to stabilize $|\phi|$ at small expectation values
because of the $D$-term constraint
\eqn\dcon{D_B=q|\phi|^2-\zeta^2=0} that has to be satisfied to
preserve supersymmetry. This forces $\phi$ to acquire a nonzero VEV
$|\phi|^2=\zeta^2/q$. Hence the $b$ has axion decay constant
\eqn\bcons{F_b=\sqrt{2}|\phi|=M_P\sqrt{k\alpha_C\Tr B\over 24\pi q}
.} From \fate,\ \bcons\ we see that $F_b\sim F_a$ so the optimistic
hypothesis that $F_b\ll F_a$ does not apply. Since the axion decays
scales of $a,b$ are comparable, to get a quantitative description,
we need to keep in the effective action both axions $a,b$ and the
anomalous gauge field $B_\mu$: \eqn\raman{-\half F_a^2 (\p_\mu a
-q_a B_\mu)^2-\half F_b^2(\p_\mu b- q_b B_\mu)^2-{1\over 4 g_B^2}\tr
F_{B,\mu\nu} F_B^{\mu\nu}+  a {1\over 16\pi^2}\tr (F_\wedge F)_\QCD}
Here $q_a={\Tr B/12}, q_b=q$. Diagonalizing \raman, it can be shown
that one linear combination of $a,b$ gets eaten by the $U(1)_B$
gauge field, giving it a string scale mass.  The other linear
combination survives to low energies as a Peccei-Quinn axion. Since
$F_a,F_b\sim M_s,$ the PQ axion has roughly string scale decay
constant \fiterms.\  Having several fields $\phi_\alpha$ charged
under $U(1)_B$ does not appear likely to change this conclusion.

\medskip\noindent{\it Anomalous $U(1)$'s in Type II String Theory}

\nref\poppitz{E. Poppitz, ``On the One Loop Fayet-Iliopoulos Term
in Chiral Four-Dimensional Type I Orbifolds,'' Nucl. Phys. {\bf
B542} 31, hep-th/9810010.} \nref\nilles{Z. Lalak, S. Lavignac and
H. P. Nilles,``String Dualities in The Presence of Anomalous U(1)
Symmetries,'' Nucl. Phys. {\bf B559} 48 (1999), hep-th/9903160.}
\nref\rabadan{L. E. Ibanez, R. Rabadan and A. M.
Uranga,``Anomalous U(1)'s in Type I and Type IIB D=4, N=1 String
Vacua,'' Nucl. Phys. {\bf B542} (1999) 112, hep-th/9808139.}
\nref\lawrence{A. Lawrence and J. McGreevy,``D-Terms and D-Strings
in Open String Models,'' JHEP 0410:056 (2004)
hep-th/0409284.}\nref\cycles{S. Kachru and J. McGreevy,
``Supersymmetric Three Cycles and Supersymmetry Breaking,'' Phys.
Rev. {\bf D61} (2000) 026001, hep-th/9908135.}\nref\douglas{M. R.
Douglas, ``$D$-Branes, Categories and ${\cal N}=1$
Supersymmetry,'' J. Math. Phys. {\bf 42} (2001) 2818,
hep-th/0011017.}

It may be possible to make models with low axion decay constant in
Type II string with intersecting $D$-branes. Here, the axion that
cancels the $U(1)_B$ anomaly is a twisted RR axion
\refs{\rabadan,\nilles}. At tree level, the FI-term is determined in
Type IIA string by complex structure moduli and in Type IIB string
by Kahler moduli \douglas.\ It has been argued that the one loop
contribution to FI-terms (that in heterotic string theory generates
string scale FI-term) is absent in Type II-orientifolds by
reinterpreting the FI terms as closed string tadpoles
\refs{\poppitz,\lawrence}. Hence, there might be a value of moduli
for which the FI-term of the anomalous $U(1)$ is well below the
string scale.

\newsec{$M$-Theory On A Manifold Of $G_2$ Holonomy}

In the present section, we consider $M$-theory on a seven-manifold
$X$ of $G_2$ holonomy, which we call $X$.  If one is not concerned
about maintaining ${\cal N}=1$ supersymmetry in four dimensions,
much of the discussion applies to $M$-theory compactification on
more general seven-manifolds. We suppose that $X$ contains a
three-manifold $Q$ of orbifold singularities, leading to
four-dimensional gauge fields. For example, $Q$ may be a locus of
${\bf A}_2$ singularities, leading to color $SU(3)$ gauge fields,
or ${\bf A}_4$ singularities, leading to a theory somewhat similar
to $SU(5)$ GUT's in four dimensions. For the qualitative
investigation of the axion coupling parameters, such details are
inessential.

Section 4 was also devoted to $M$-theory on a seven-manifold, of
the form $Z\times I$.  The main difference is the strategy for
getting Standard Model gauge fields; in section 4, these were
supposed to arise at the boundary of the world, while in our
present discussion, we will get gauge fields from orbifold
singularities.

{}From \olygo, it follows that in reduction on $M\times X$, the
four-dimensional Planck scale is given by \eqn\tolly{M_P^2={4\pi
V_X\over \ell_{11}^9}.} The action for $SU(3)$ (or $SU(5)$) gauge
fields along $M\times Q$ is\foot{To get this formula, start with
eqn. (13.3.25) of \polchinski, which shows that the kinetic energy
for gauge fields on a Type IIA $D6$-brane is $(1/8\pi
g_s\ell_s^3)\int d^7x\sqrt{-g}\,{\bf tr}\,F_{\mu\nu}F^{\mu\nu}$.
Then convert to $M$-theory parameters via the relation
$g_s\ell_s^3=\ell_{11}^3$, which follows from the equality of
$D2$-brane and $M2$-brane tensions using our conventions in
section 2.} \eqn\uctu{{1\over 8\pi\ell_{11}^3}\int_{M\times Q}d^7x
\sqrt{-g}\,{\bf tr}\,F_{\mu\nu}F^{\mu\nu}.} Upon reduction to four
dimensions and recalling the convention ${\bf tr}\,t^at^b={1\over
2}\delta^{ab}$ for Lie algebra generators, this becomes
\eqn\nuctu{{V_Q\over 16\pi\ell_{11}^3}\int_M d^4x\sqrt{-g}
F_{\mu\nu}^aF^{\mu\nu\,a}.} So the color $SU(3)$ gauge coupling at
the compactification scale obeys $g_C^2=4\pi\ell_{11}^3/V_Q$ or
equivalently \eqn\cuvu{\alpha_C={\ell_{11}^3\over V_Q}.}

Axions arise from zero modes of the three-form field $C$ of
$M$-theory.  If $\gamma_i$, $i=1,\dots, b_3(X)$ are the harmonic
three-forms on $X$, normalized so that
\eqn\tolnito{\int_{D_i}\gamma_j=\delta_{ij},} for a suitable basis
of three-cycles $D_i$, then we make an ansatz \eqn\invi{C={1\over
2\pi}\sum_i c_i \gamma_i,} where $c_i$ are massless fields on $M$.
The kinetic energy for the $c_i$ is obtained by dimensional
reduction from \olygo.  It is \eqn\yolno{{1\over
2\pi\ell_{11}^3}\int_M {1\over 2}\tilde\gamma^M_{ij}dc_i\wedge
\star dc_j,} where \eqn\colno{\tilde\gamma^M_{ij}=\int_X
\gamma_i\wedge \star\gamma_j.}

The $c_i$  have approximate shift symmetries because of the
underlying gauge-invariance of the $C$-field.  They have
axion-like couplings because the $M$-theory effective action at an
orbifold singularity includes a coupling
\eqn\bilno{2\pi\int_{M\times Q}C\wedge{1\over 8\pi^2}{\bf tr}\,
F\wedge F.} The existence and coefficient of this coupling follow
from the analogous coupling on the world-volume of a $D6$-brane in
Type IIA superstring theory; note that the coupling in \bilno\ is
quantized for topological reasons, so it is completely determined
in $M$-theory by what it is in Type IIA.  \bilno\ reduces in four
dimensions to \eqn\hilno{\sum_ik_i\int_M c_i{1\over 8\pi^2}{\bf
tr}\,F\wedge F,} with \eqn\logiko{k_i=\int_Q \gamma_i.} Thus, the
modes $c_i$ indeed have the couplings of axions.

Let us define a ``radius'' $R$ of $X$ by $V_X=R^7$.  Under an
overall scaling of the metric of $X$, the integral in \yolno\ scales
like $R$.  Thus if $X$ is reasonably isotropic, a ``generic'' linear
combination of the $c_i$ has \eqn\milko{F_c^2={xR\over
2\pi\ell_{11}^3},} where $x$ is formally of order 1. Let us first
assume that $Q$ is a generic three-cycle, say with $R_Q$ of order
$R$. If we set $R_Q=R$, then \milko\ and \cuvu\ give us
$F_c=x\alpha_CM_P/2\pi\sqrt 2$, a familiar sort of formula. With
$R_Q$ of order $R$, compactification on a manifold of $G_2$ holonomy
can give reasonable GUT-like phenomenology and a sensible relations
between Newton's constant and the GUT scale; these issues have been
explored in \ref\fw{T. Friedmann and E. Witten, ``Unification Scale,
Proton Decay, And Manifolds Of $G_2$ Holonomy,'' Adv. Theor. Math.
Phys. {\bf 7} (2003) 577.}. In this case, evidently, we expect the
axion coupling parameter to be close to the GUT scale.  The action
for a membrane wrapped on $Q$ is precisely $I=2\pi
R_Q^3/\ell_{11}^3=2\pi/\alpha_C$, and indeed such a membrane is
equivalent to an instanton of the gauge theory. For $R_Q\sim R$,
generic membrane instantons have actions comparable to this. Hence,
with $\alpha_C\sim 1/25$, explicit PQ violation at high energies
might be small enough, subject to the usual caveats, to lead to a
solution of the strong CP problem.

Alternatively, we can ask what are the necessary parameters that
give $F_c$ in the range allowed by the usual cosmological arguments.
We can solve \tolly\ and \milko\ for $\ell_{11}$ and $R$ in terms of
$F_c$ and $M_P$: \eqn\hobo{\eqalign{\ell_{11} & = {1\over
F_c}\left(M_P\over F_c\right)^{1/6}{x^{7/12}\over
2^{3/4}\pi^{2/3}}\cr R& = \ell_{11}\left({M_P\over
F_c}\right)^{1/3}{1\over 2^{1/2}\pi^{1/3}}\cr}.}

For $10^{9}\,{\rm GeV}\lesssim F_c\lesssim 10^{12}\,{\rm GeV}$, and
taking $x=1$, we have roughly \eqn\oflog{\eqalign{{10\over
F_c}&\gtrsim \ell_{11}\gtrsim {3\over F_c}\cr 500\ell_{11}&\grtsim
R\grtsim 50\ell_{11}. \cr}} Even at the lower end of this range, a
generic membrane instanton action $2\pi R^3/\ell_{11}^3$ is
prohibitively large, so significant PQ violation will come only from
gauge instantons -- or from membranes wrapped on the vanishing
cycles that we discuss later.

Similarly, we define a ``radius'' $R_Q$ of $Q$ such that
$V_Q=R_Q^3$.  From \cuvu, $\alpha_C=\ell_{11}^3/R_Q^3$.  To get a
reasonable value of $\alpha_C$, $R_Q$ must be fairly close to
$\ell_{11}$. It follows, therefore, from \oflog\ that if $F_c$ is
to be in the usual cosmological range, $R_Q$ must be substantially
less than $R$.  So the three-cycle on which gauge fields are
supported must be substantially smaller than a generic three-cycle
in $X$.

It is believed to be possible for a manifold $X$ of $G_2$ holonomy
to develop, as its moduli are varied, a supersymmetric ``vanishing
cycles'' -- a three cycle  that collapse to zero even as the metric
on the rest of $X$ has a limit.  The example about which most is
known is the case that the vanishing cycle is a three-sphere $S^3$
-- or an orbifold quotient thereof, $S^3/\Gamma$, with $\Gamma $ a
finite group of symmetries.  The local structure near the vanishing
cycle is given by an explicitly known \lref\bryant{R. Bryant, S.
Salamon, ``On the Construction of Some Complete Metrics with
Exceptional Holonomy,'' Duke Math. J. {\bf 58} (1989) 829.}
\lref\gibbons{G. W. Gibbons, D. N. Page, C. N. Pope, ``Einstein
Metrics on $S^3$, $R^3$ and $R^4$ Bundles,'' Comm. Math. Physics
{\bf 127} (1990) 529.} \refs{\bryant, \gibbons} asymptotically
conical metric of $G_2$ holonomy on the manifold $S^3\times
\Bbb{R}^4$ (or an orbifold quotient thereof). Its role in $M$-theory
has been analyzed in some detail \nref\vag{M. F. Atiyah, J. M.
Maldacena, and C. Vafa, ``An $M$ Theory Flop As A
Large $N$ Duality,'' J. Math. Phys. {\bf 42} (2001) 3209, hep-th/0011256.}%
\nref\aw{M. F. Atiyah and E. Witten, ``$M$ Theory Dynamics On A Manifold Of
$G_2$ Holonomy,'' Adv. Theor. Math. Phys. {\bf 6} (2003) 1, hep-th/0107177.}%
\nref\tf{Tamar Friedmann, ``On The Quantum Moduli Space Of $M$
Theory Compactifications,'' Nucl. Phys. {\bf B635} (2002) 384.}%
 \refs{\vag - \tf}.

In our problem, to get $F_c$ in the usual cosmological range, it
is natural to assume that $Q$ is such a vanishing cycle -- and the
vacuum is near a point in moduli space at which $V_Q$ would go to
zero.  (We do not know why the vacuum would be near such a point,
but we recall that for Type IIB superstring theory, mechanisms
have been proposed \ref\who{S. B. Giddings, S. Kachru, and J.
Polchinski, ``Hierarchies From Fluxes In String
Compactifications,'' Phys. Rev. {\bf D66} (2002) 106006,
hep-th/0105097.} that can lead to a vacuum near a point with a
vanishing cycle.) In fact, the example that the local structure is
an orbifold quotient of $S^3\times \Bbb{R}^4$ has some of the
right properties for us. By dividing by a group
$\Gamma'=\Bbb{Z}_3$ or $\Gamma'=\Bbb{Z}_5$ that acts only on
$\Bbb{R}^4$ with an isolated fixed point at the origin, one can
get $SU(3)$ or $SU(5)$ gauge fields supported on $S^3$. In the
$SU(5)$ case, if one also divides by a group $\Gamma$ that acts
freely on $S^3$, one introduces the possibility of spontaneously
breaking $SU(5)$ to the Standard Model.  To include quarks and
leptons, one would need to complicate the singularity structure
\ref\wa{B. Acharya and E. Witten, ``Chiral Fermions From Manifolds
Of $G_2$ Holonomy,'' hep-th/0109152.}. Such models have some
obvious potential problems; because the compactification scale is
well below the usual GUT scale (to get $F_c$ in the usual
cosmological range) it will be hard to avoid rapid proton decay or
to get the right low energy gauge couplings. A more complicated
structure of vanishing cycles might be necessary.

\bigskip\noindent{\it Axion Physics With A Small $S^3$}

In the case when $X$ develops a vanishing $S^3$, we can be more
precise in estimating $F_c$, as the explicit metric of the local
$G_2$ holonomy manifold with vanishing $S^3$ is known \gibbons.\
Locally the manifold has the topology $S^3\times \Bbb{R}^4/\Gamma,$
where $\Gamma$ is a finite subgroup of $SU(2)$ with which we
orbifold $\Bbb{R}^4$ to get nonabelian gauge symmetry.
Asymptotically, the manifold is a cone over $S^3\times S^3/\Gamma$
where the $S^3$ is homologous to the vanishing three-cycle $Q$ and
$S^3/\Gamma$ is the quotient of the unit three-sphere in
$\Bbb{R}^4.$

The axion decay constant is  \yolno\ \eqn\apop{F_c^2={1\over 2\pi
\ell_{11}^3}\int_X \gamma_Q\wedge\star\gamma_Q,} so a more precise
evaluation of $F_c$ amounts to a calculation of the norm of
$\gamma_Q$.

For $F_c$ in the range allowed by standard cosmological arguments,
the estimate \oflog\  gave $R>>R_Q$.  Hence, to a high degree of
accuracy, we can neglect the finite size of $Q$ and treat $X$
locally as a cone over $S^3\times S^3/\Gamma$. We work in the
``upstairs picture'' on a cone over $S^3\times S^3.$ To take into
account quotienting by $\Gamma,$ one divides $F^2_c$ at the end of
the calculation by $|\Gamma|$.

To write down the metric explicitly, we introduce invariant
one-forms $\sigma_i,\Sigma_i, ~ i=1,\dots,3$ on the two
three-spheres. We normalize them so that the usual round metric on a
three-sphere with radius one is $ds^2=\sum_i \sigma_i^2$. With this
normalization, $\sigma_i$ obey the usual $SU(2)$ relations
$d\sigma_i=-\half \epsilon_{ijk} \sigma_j\wedge \sigma_k$ and the
volume form on $S^3$ is just $\sigma_1\wedge\sigma_2\wedge\sigma_3.$
The metric on the base of the cone, $S^3\times S^3$, is the squashed
Einstein metric \eqn\squa{\eqalign{ds^2&=dr^2+ {r^2\over 9}
(\sigma_i-{\Sigma_i\over 2})^2+{r^2\over 12} \Sigma_i^2,\cr &\equiv
dr^2+\nu_i^2+e_i^2.}} In the second line we introduced the
orthonormal basis $\nu_i= r(\sigma_i-{\Sigma_i/2})/3,~
e_i=r\Sigma_i/\sqrt{12}.$ In writing the metric we chose $\Sigma_i$
to be the one-forms of the non-contractible sphere $Q$. The
vanishing cycle $Q$ is Poincare dual to a harmonic three-form
\lref\Cvetica{
  M.~Cvetic, H.~Lu and C.~N.~Pope,
  ``Brane resolution through transgression,''
  Nucl.\ Phys.\ B {\bf 600} (2001) 103,
  hep-th/0011023.
} \lref\Cveticb{
  M.~Cvetic, G.~W.~Gibbons, H.~Lu and C.~N.~Pope,
  ``Supersymmetric non-singular fractional D2-branes and NS-NS 2-branes,''
  Nucl.\ Phys.\ B {\bf 606} (2001) 18,
  hep-th/0101096.
} \refs{\Cveticb}, eq. (2.44),  which in the conical limit becomes
\eqn\thref{\omega_3={1\over r^3}
\epsilon^{ijk}\nu_i\wedge\nu_j\wedge e_k+{3\over r^3} e_1\wedge
e_2\wedge e_3.} The Hodge dual of $\omega_3$ is
\eqn\hodo{\star\omega_3={dr\over r^3}\left(\epsilon_{ijk}\nu_i\wedge
e_j\wedge e_k-3\nu_1\wedge\nu_2\wedge\nu_3\right).}

The flux of $\omega_3$ through the non-contractible $S^3$ is
\eqn\flun{\int_{S^3}\omega_3={\pi^2\over 4\sqrt{3}},} so the
three-form with unit flux through $Q$ is $\gamma_Q=4\sqrt{3}\omega_3
/\pi^2$ and the axion comes from the ansatz $C=\gamma_Q a/2\pi.$ We
estimate  the axion decay constant by substituting the explicit form
of $\gamma_Q$ into \apop\ \eqn\ase{\eqalign{F_c^2&={1\over 2\pi
\ell_{11}^3 N}\left(4\sqrt{3}\over \pi^2\right)^2{12\over
(6\sqrt{3})^3}\int_X dr\wedge d\sigma_1\wedge d\sigma_2\wedge
d\sigma_3\wedge d\Sigma_1\wedge d\Sigma_2\wedge d\Sigma_3,\cr
&={R\over \ell_{11}}{2^4 x\over 3^{5/2}\pi N} M_{11}^2,}} where $x$
is of order one and $R=V_X^{1/7}$ is the linear size of $X$ (at
which we cut off the integral). We divided the integral by
$|\Gamma|=N$ to take into account the orbifolding of  $X$ by
$\Gamma$. For $SU(5)$ gauge symmetry along $M\times Q$ we take
$\Gamma=\Bbb{Z}_5.$ With $|\Gamma|=5$, \ase\ comes out to be roughly
the same size as \milko.\ There is an additional contribution to
$F_c$ from the region that compactifies the cone. This gives an
additive contribution to $F_c^2$ of size given by \milko,\ hence it
does not change \ase\ significantly.

With the estimate \ase\ for  $F_c,$ we re-derive the values for
$M$-theory scale $M_{11}$ and the size $R$ of $X$ in terms of $M_P$
and $F_c$ \eqn\ners{\eqalign{M_{11}&=F_c\left(F_c\over
M_P\right)^{1/6}{3^{35/24}\pi^{2/3}N^{7/12}\over
2^{13/6}x^{7/12}},\cr R&=\ell_{11}\left(M_P\over
F_c\right)^{1/3}{2^{1/3}x^{1/6}\over 3^{5/12}\pi^{1/3}N^{1/6}}.}}
For $F_c$ in the range allowed by standard cosmological constraints,
setting $x=1, N=5$, we find \eqn\finra{\eqalign{{F_c\over 6}\lesssim
&M_{11}\lesssim {F_c\over 2},\cr 550\ell_{11}\gtrsim &R\gtrsim
55\ell_{11}.}} We see that the more precise calculation of $F_c$
does not change the result \oflog\ significantly.

\bigskip\noindent{\it Anisotropic Seven-Manifolds}

So far we have assumed that our manifold $X$ of $G_2$ holonomy is
reasonably isotropic, with gauge fields supported either on a
generic cycle or on a vanishing cycle.  It is also possible for
$X$ to be highly anisotropic.

A $G_2$ manifold can have two types of supersymmetric fibration: a
fibration by three-tori over a four-dimensional base, or by K3
surfaces over a three-dimensional base.  Apart from special
constructions involving orbifolds (as opposed to more generic
$G_2$ manifolds), these are the most obvious kinds of highly
anisotropic $G_2$ manifolds.  Much of our discussion, in any case,
has nothing to do with supersymmetry and would carry over to other
kinds of highly anistopropic manifold.

Some features of these two types of example can be treated
together. We refer to the fiber as $F$ and the base as $B$; we
write $d$ for the dimension of the fiber (so $d=3$ or 4), and we
express the volumes as $V_X=V_FV_B$, $V_F=R_F^d$, $V_B=R_B^{7-d}$.
We only need to consider the case that $R_F<<R_B$. If $R_F\sim
R_B$, we are back in the case we have already considered of a more
or less isotropic manifold.  It is generically impossible in a
supersymmetric fibration to have $R_F>>R_B$, but in any case, when
this occurs one should look for an alternative description with
the roles of $F$ and $B$ exchanged.

The four-dimensional reduced Planck mass obviously becomes
\eqn\yugop{M_P^2={4\pi V_FV_B\over \ell_{11}^9}.} The
generalization of the formula \cuvu\ for $\alpha_C$ depends on the
geometry.  Let us suppose that $Q$ has $a$ dimensions wrapped on
$F$ and $3-a$ wrapped on $B$, and that $F$ and $B$ are each more
or less anisotropic.  In this case,
\eqn\tupgo{\alpha_C={\ell_{11}^3\over V_Q}=y{\ell_{11}^3\over
R_F^aR_B^{3-a}}} where $y$ is a constant of order 1.  Obviously,
this formula could be substantially changed if $Q$ wraps vanishing
cycles in either $F$ or $B$.  This would lead to a hybrid of our
present discussion of anisotropic $X$ with the earlier discussion
of vanishing cycles.

Similarly, the estimation of $F_c$ depends on what kind of
three-cycle we consider.  If $\omega$ is a harmonic three-form on
$X$ with $b$ indices tangent to $F$ and $3-b$ indices tangent, and
we keep the topological class of $\omega$ fixed while letting
$R_F$ and $R_B$ vary, then $\int_X\omega\wedge \star \omega \sim
V_X/R_F^{2b}R_B^{6-2b}$.  In general (as described most fully by
the appropriate ``spectral sequence''), a harmonic form is a sum
of components with different values of $b$.  For $R_F<<R_B$, the
component with the largest $b$ will dominate. The analog of
\milko\ for a generic axion $c$ with a given largest value of $b$
is \eqn\olbo{F_c^2={xV_X\over 2\pi\ell_{11}^3R_F^{2b}R_B^{6-2b}}.}

To make $F_c$ small, we want $b$ to be as small as possible.
However, if $b<a$, then the mode $c$ does not have an axionic
coupling to gauge fields that are supported on $M\times Q$.  The
smallest relevant $F_c$ therefore has $b=a$.  For such a mode, we
can combine \olbo, \tupgo, and \yugop\ to the familiar order of
magnitude relation $F_c\sim \alpha_CM_P$.

Thus, taking $X$ to be highly anisotropic is not an efficient way
to reduce the order of magnitude of $F_c$.  To get a small $F_c$
from a fibration with $R_F<<R_B$, we must assume that $Q$ is
wrapped on a vanishing cycle in either $F$ or $B$.

The arguments that we have just given also apply to other examples
(such as gauge fields supported on Type II $D$-branes) that we
will consider later.

\newsec{Intersecting $D$-brane Models}

In this section, we consider intersecting $D$-brane models both in
type IIA and IIB string theory. We assume that gauge symmetry lives
on $D(3+q)$-branes which are extended along the four noncompact
dimensions and wrap a $q$-cycle $Q$ in the compactification
manifold. In Type IIA, one takes a stack of five $D6$-branes wrapped
around $M\times Q$ where $M$ is the Minkowski space and $Q$ is a
compact special Lagrangian three-cycle in the compact manifold $X$.
In Type IIB, one takes instead $D3$, $D5$ or $D7$-branes wrapping a
holomorphic cycle in $X$. Three $D$-branes wrapping a cycle $Q$
support $U(3)$ gauge theory whose nonabelian part could be the QCD
gauge symmetry. Five $D$-branes would lead instead to a theory
somewhat similar to an $SU(5)$ GUT. In case that the $q$-cycle $Q$
has nonzero $\pi_1(Q),$ we can break the GUT gauge group down to the
Standard Model gauge group by turning on discrete Wilson lines.

The low energy effective supergravity action contains the
gravitational term \molmo\ \eqn\grterm{S_{GR}={2\pi\over g_s^2
\ell_s^8}\int \sqrt{-g}R.} Dimensionally reducing the Einstein
action to four dimensions determines the Planck mass
\eqn\aplanck{M_P^2={4\pi{V_X\over g_s^2 \ell_s^8}}.} The low energy
action of the RR $q$-form field $C_q$ is, \polchinski\ eq. (13.3.5),
\eqn\reac{-{2\pi\over \ell_s^{8-2q} }\int \half F_{q+1}\wedge\star
F_{q+1}+2\pi\int C_{q},} where the second term is the coupling of
the $q$ form to $D(q-1)$ branes. We normalized $C_q$ so that the RR
field strength $F_{q+1}$ has integer periods. Thus, our RR-field is
related to the usual one $C_{q,conv}$ by $C_{q,conv}=\ell_s^q C_q.$

The effective action of the gauge theory living on the $D$-branes
is, \polchinski\ eq.  (13.3.25) \eqn\sixe{S_{Y\negthinspace
M}=-{1\over 4(2\pi)g_s \ell_s^{4+q}}\int d^qx\sqrt{-g}\, {\bf tr}
F_{\mu\nu}F^{\mu\nu},} where the trace is in the fundamental ${\bf
N}$ representation of $SU(N)$.  Reducing the gauge action on $Q$ to
four dimensions leads to the action \eqn\yma{-{V_Q\over 8 (2\pi)
g_s\ell_s^q}\int d^4x \sqrt{-g} F^a_{\mu\nu}F^{\mu\nu~a},} where we
used the normalization of $SU(N)$ generators ${\bf tr} \,t^a
t^b=\half \delta^{ab}$ that is conventional in the GUT literature.
>From \yma\ we read off the four-dimensional gauge coupling
\eqn\aguit{\alpha_{GUT}={g_s\ell_s^q\over V_Q}.}

In type II string theory, the axions come from zero modes of the
$q$-form gauge field $C_q.$ The axions are the phases $2\pi\int_{Q}
C_q$ necessary for the complete definition of the string theory path
integral in the presence of $D(q-1)$ branes. They are angular
variables with period $2\pi.$  Let $Q_\alpha,\, \alpha=1,\dots,
b_q(X)$ be an integral basis of the homology group $H_q(X,\Bbb{Z})$
modulo torsion. We take $\omega_\alpha$ to be harmonic
representatives of the basis of $H^q(X)$ dual to the basis
$Q_\alpha,$ so that $\int_{Q_\alpha}
\omega_\beta=\delta_{\alpha\beta}.$

The axions come from the ansatz
\eqn\exmex{C_q={1\over2\pi}\sum_\alpha a_\alpha \omega_\alpha,\qquad
\alpha=1,\dots, b_q(X).} We included a factor of $1/2\pi$ so that
$a_\alpha$ have period $2\pi.$ Substituting this into the RR-field
effective action \reac\ we get the kinetic energy of the axions
\eqn\remex{S=-\half\sum_{\alpha,\beta} \gamma_{\a\b} \p_\m a_\a
\p^\m a_\b,} where \eqn\gemex{\g_{\a\b}={1\over 2\pi
\ell_s^{8-2q}}\int_X \om_\a\wedge \star \omega_\b.} The axions
acquire axionic couplings from the D-brane Chern-Simons term,
(13.3.18) of \polchinski\ \eqn\aco{2\pi\int_{M\times Q} C_q \wedge
{1\over 8\pi^2}{\bf tr}\, F\wedge F .} Dimensionally reducing this
to four dimensions using the ansatz \exmex\ leads to the couplings
\eqn\acoupling{\sum_\alpha r_\alpha \int a_\alpha {{\bf tr} F\wedge
F \over 8\pi^2},} where $r_\alpha=\int_Q \omega_\alpha$ are
integers.

Let us first consider branes wrapping a $q$-cycle $Q$ with $q>0.$
The case $q=0$ with a stack of $D3$ branes localized at a point will
be considered later. We define $R$ to be linear size of $X$, so that
$V_X=R^6.$ In terms of $R$, the Planck mass is
\eqn\nonna{M_P^2={4\pi R^6\over g_s^2\ls^8}.} For a generic axion,
$\int_X \omega_Q\wedge \star\omega_Q=xR^{6-2q},$ where $x$ is of
order one, so \eqn\genericax{F_a=\sqrt{x R^{6-2q}\over 2\pi
\ls^{8-2q}}=M_P\left(\ell_s\over R\right)^q\sqrt{{x g_s^2\over
8\pi^2}},} where $x$ is a dimensionless number of order one. Hence
$F_a$ is in the phenomenologically preferred range if either
$g_s<<1$ or $R>>\ell_s.$  The first possibility is excluded by
considerations about gauge couplings. Indeed, if the gauge symmetry
comes from D-branes wrapping a $q$-cycle $Q$ of radius
$R_Q=V_Q^{1/q}$, the four-dimensional gauge coupling is
$\alpha_C={g_s \ell_s^q/R_Q^q.}$ To have a good perturbative
description, $R_Q$ should not be much smaller than $\ls$ so that
$\alpha'$ corrections are suppressed (otherwise we can go into a
T-dual description, in which $R_Q\gtrsim \ell_s$. Setting
$R_Q\gtrsim\ell_s$ gives an upper bound on the string coupling
$g_s\gtrsim \alpha_C.$ Hence, we do not have a freedom to lower
$g_s$ to arbitrarily small values. The latter possibility leads to
low scale axions if the compactification manifold has very large
size in string units.

To estimate the parameters of the compactification that lead to
phenomenologically preferred axion decay constants, we express $R$
and $M_s=\ls^{-1}$ in terms of $F_a,M_P$  from \nonna\ and
\genericax:\ \eqn\etex{\eqalign{R&=\ls \l(M_P\over F_a\r)^{1\over
q}\l(x g_s^2\over 8\pi^2\r)^{1\over 2q},\cr M_s&=F_a \l(F_a\over
M_P\r)^{3-q\over q}\l(2\pi\over x\r)^{1\over 2}\l(8\pi^2\over
xg_s^2\r)^{3-q\over 2q}
.}} Requiring that the axion decay constant falls into the range
$10^9\GeV\leq F_a\leq 10^{12}\GeV$ implies large compactification
radius $R>>\ell_s$ and a low string scale $M_s<<M_{GUT}.$

In these compactifications, gauge symmetry lives on D-branes
wrapping a cycle $Q$. The radius of $Q$ is at most a few string
lengths; otherwise, the string coupling necessary for getting the
correct four-dimensional gauge coupling $g_s=\alpha_C
(R/\ell_s)^q$ get nonperturbatively large and our target space
effective description breaks down. Hence, $Q$ is a `vanishing'
cycle in $X$ with $R_Q<<R$. A possible moduli stabilization
mechanism that fixes the moduli of $X$ in this regime has been
recently discussed in \refs{\BalasubramanianZX, \ConlonTQ}. The
actual results for the preferred compactification parameters
depend on which $q$-form RR-field does the axion originate from
and on the geometry of the compactification manifold. In section
7.1, we give a more precise treatment of compactifications with
vanishing cycles in two concrete examples.

\medskip\noindent{\it Asymmetric Calabi-Yau Manifolds}

As an alternative to assuming that $Q$ is a vanishing cycle, one can
ask whether taking an asymmetric Calabi-Yau manifold that is a
fibration with small fiber $F$ over a large base $B$ could
substantially lower the axion coupling parameter.  Based on our
experience with heterotic string in section 3 we expect that the
decay constants of axions in asymmetric compactifications are still
around the GUT scale.

To explore this question, we will consider supersymmetric fibrations
with $T^2, T^3$ or $K3$ fibers, but our conclusions apply to more
general fibrations (including nonsupersymmetric ones).  The volume
of the Calabi-Yau manifold is $V_X=V_BV_F,$ hence the reduced Planck
mass \aplanck\ is \eqn\redumes{M_P^2={4\pi V_B V_F\over g_s^2
\ls^8}.} We let $R_F$ be the `radius' of the fibre, so that
$V_F=R_F^d$ and $R_B$ be the `radius' of the base $V_B=R_B^{6-d}.$
An axion coming from a zero mode of the $q$-form field $C_q$ with
$b$ indices along the fibre and $q-b$ indices along the base has
axion coupling parameter \gemex\ \eqn\horcux{F_a^2={xV_X\over 2\pi
\ls^{8-2q} R_B^{2q-2b} R_F^{2b}},} where $x$ is a dimensionless
number of order one. We assume that the gauge symmetry comes from a
$D(3+q)$-branes wrapping a cycle with $a$ dimensions wrapped around
the fibre and $q-a$ dimensions wrapped along the base, so the gauge
coupling is \eqn\aniga{\alpha_C={g_s \ls^q\over R_F^a R_B^{q-a}}.}
The axionic coupling is nonzero only for axions with $b=a.$ For
these, we find \eqn\asymtype{F_a={\sqrt{x}\alpha_C\over
2\pi}{M_P\over \sqrt{2}},} which gives the familiar answer  $\sim
10^{16}\GeV.$

\vskip .3 in \noindent{\it $D3$-Brane Models}

The cycle that the D3-branes ``wrap'' is a point which results in a
different behavior of the axion decay parameter compared to other
string theories. Hence, there is no hierarchy between the size of
the vanishing cycle and the size of the compactification manifold
that could help lower the axion decay constant.  The low energy
gauge group on $N$ $D3$-branes at a generic point in $X$ is $U(N).$
The gauge coupling is fixed by the string coupling \sixe:
\eqn\acgs{\alpha_C=g_s.}

The axions are four-dimensional fields coming from reduction of the
RR zero-form. A harmonic zero-form is just a constant, so we use the
ansatz \eqn\cnota{C_0={a\over 2\pi},} where $a$ is a
four-dimensional pseudo-scalar field. It follows from the D-brane
Chern-Simons coupling \aco\ that the axion has $r=1$ coupling to the
$QCD$ instanton density \eqn\cnoco{\int a {{\bf tr} F\wedge F\over
8\pi^2}.} The kinetic energy of the RR zero-form \reac\ is easily
reduced to four dimensions, giving the axion kinetic energy
\eqn\tfor{{V_X\over 2\pi \ls^8}\int d^4x\left( -\half \p_\m a\p^\m
a\right),} whence the axion coupling constant is
\eqn\threeax{F_a=\sqrt{V_X\over2\pi
\ls^8}={\alpha_C\over2\pi}{M_P\over \sqrt{2}}.} If we take
$\alpha_C\sim 1/25,$ we get $F_a=1.1\times 10^{16}\GeV,$ which is
the same as the axion coupling parameter of the model-independent
axion in weakly coupled heterotic string theory.  The shift symmetry
of the axion is explicitly broken by $D(-1)$-brane instantons that
are located on the $D3$-brane worldvolume. These instantons are
equivalent the $SU(N)$ gauge theory instantons. Their action is
$I=2\pi/g_s=2\pi/\alpha_C.$ With $\alpha_C\sim 1/25,$ their action
is $I\sim 157$, so the explicit violation of the shift symmetry
might be small enough for the axion to be a candidate for
Peccey-Quinn axion.

\subsec{Intersecting $D$-brane Models With Small Cycles}

In the previous subsection we found that the axion coupling
parameter can be lowered in type II string theory, if the the gauge
symmetry comes from $D$-branes wrapping a vanishing cycle. To lower
$F_a$ into the range $10^9\GeV<F_a<10^{12}\GeV$, the radius of the
compactification manifold has to be much larger than the string
length. This lowers the string scale relative to the Planck scale.
We estimated this in \etex.\

To get a more precise estimate of the physical scales involved in
getting $F_a$ in the phenomenologically preferred range, we study in
detail the compactification of Type II string on a CY manifold $X$
that is developing a conifold singularity. The vanishing cycle at
the tip is either an $S^2$ or an $S^3$, depending on whether the
conifold is resolved or deformed. To get $SU(5)$ gauge symmetry, we
wrap a stack of five $D5$ or $D6$-branes around the vanishing cycle.
The $D$-branes warp the geometry in a region of size $\ell\sim (g_s
N)^{1/4} \ell_s$, as is familiar from AdS/CFT correspondence. For
$N=5,$ that is necessary for $SU(5)$ gauge symmetry, the warped
region has size around the string length. Since we took the radius
of the CY manifold much larger than the string length, we can
neglect the effect of the warping on the axion coupling parameter
$F_a\sim \int_X \omega\wedge \star\omega$.

\nref\candelas{P. Candelas and X. C. de la Ossa,``Comments On
Conifolds,'' Nucl. Phys. B {\bf 342}, 246 (1990).} \nref\herzog{C.
P. Herzog, I. R. Klebanov and P. Ouyang, ``Remarks on the Warped
Deformed Conifold,'' arXiv:hep-th/0108101.}

The conifold is a cone over $T^{1,1}$, where $T^{1,1}$ is
topologically $S^2\times S^3.$ It is an $S^1$ fibration over
$S^2\times S^2$. Its the metric is \refs{\candelas,\herzog}
\eqn\conm{ds^2=dr^2+r^2 ds_{T^{1,1}}.} To describe the metric of
$T^{1,1}$, we parametrize the $S^1$ fiber with $\psi$, which ranges
from 0 to $4\pi$ and the two $S^2$'s with spherical coordinates
$(\theta_i,\phi_i), i=1,2$. We introduce the following basis of
one-forms \herzog\ \eqn\onef{\eqalign{g^1&={e^1-e^3\over
\sqrt{2}},\quad g^2={e^2-e^4\over \sqrt{2}},\cr g^3&={e^1+e^3\over
\sqrt{2}},\quad g^4={e^2+e^4\over\sqrt{2}},\cr g^5&=e^5,}} where
\eqn\eones{\eqalign{e^1&=-\sin\theta_1 d\phi_1,\quad e^2=d\theta_1,
\cr e^3&=\cos\psi\sin\theta_2d\phi_2-\sin\psi d\theta_2,\cr
e^4&=\sin\psi\sin\theta_2d\phi_2+\cos\psi d\theta_2,\cr
e^5&=d\psi+\cos\theta_1d\phi_1+\cos\theta_2d\phi_2.}} In terms of
these, the $T^{1,1}$ metric takes the form
\eqn\tmetric{ds_{T^{1,1}}={1\over 9} (g^5)^2+{1\over 6} \sum_{i=1}^4
(g^i)^2.} On the conifold, there are harmonic two- and three-forms
\eqn\clof{\eqalign{\omega_2&=\half(g^1\wedge g^2+g^3\wedge
g^4)
,\cr \omega_3&=g^5\wedge \omega_2.}} Their Hodge duals are
\eqn\hobe{\star \omega_2 ={r\over 3} dr\wedge \omega_3\quad \star
\omega_3=-3{dr \over r}\wedge \omega_2.} If we think  of $T^{1,1}$
as an $S^2$ fibration over $S^3,$ then $\omega_2$ has nonzero flux
through the $S^2$ fiber and $\omega_3$ has nonzero flux through the
$S^3$ base of the fibration. To find these fluxes, we take a
representative $S^2$ fiber with $\psi=0, \theta_1=\theta_2$ and
$\phi_1=-\phi_2$. The $S^3$ base can be defined with the equations
$\psi_2=\phi_2=0.$ Integrating the explicit expressions \clof\ for
the harmonic two and three-forms over the cycle representatives
gives \eqn\normal{\int_{S^2} \omega_2=4\pi, \quad\int_{S^3}
\omega_3=8\pi^2.}

The use of these formulas is the following. The cone over
$T^{1,1}$ can be slightly resolved or deformed to make a smooth
six-manifold $X'$ with a small $S^2$ or $S^3$ at its center.  (We
think of $X'$ as an approximation to part of a compact Calabi-Yau
manifold $X$.) We obtain gauge theory by wrapping $D$-branes on
the vanishing cycle, that is, on the small $S^2$ or $S^3$.  The
gauge coupling is inversely proportional to the volume of the
vanishing cycle, and so depends crucially on the details of the
resolution or deformation of the cone.  The axion that couples to
these gauge fields comes from a harmonic two-form or three-form on
$X$, which we approximate by a harmonic two-form on $X'$.  As
there is no ${\Bbb L}^2$ harmonic two-form or three-form on $X'$,
this form is not supported near the vanishing cycle, and in
describing it we can simply approximate $X'$ by the cone.  The
relevant harmonic two-form and three-form on the cone are simply
the pullbacks of the harmonic forms $\omega_2$ and $\omega_3$ on
$T^{1,1}$.

\medskip\noindent{\it Type IIB}

Let us first consider the type IIB string theory  with gauge
symmetry coming from $D5$-branes wrapping the vanishing $S^2.$  If
the radius of the $S^2$ is $r_0,$ the gauge coupling is
\eqn\gace{\alpha_{GUT}={g_s \ell_s^2\over 4\pi r_0^2}.}  The axion
comes from a mode of the RR two-form field $C_2$ that is the product
of a four-dimensional field $a$ with a harmonic  two-form field that
has nonzero flux over the vanishing cycle $S^2$.  Since this
harmonic two-form is not supported near the vanishing cycle, we can
approximate it by  a harmonic form on the cone, which in fact is a
pullback of the harmonic form $\omega_2$ from the five-manifold
$T^{1,1}$.  Thus, the ansatz for the RR two-form is
\eqn\flata{C_2={\omega_2\over 4\pi}{a\over 2\pi}.} We used \normal\
to normalize $a$ to have $2\pi$ periods. To find the axion decay
constant, we substitute the ansatz \flata\ into the formula for the
axion decay constant \gemex.\ With the help of
\hobe,\ we get \eqn\inb{F_b^2={1\over 3(32\pi^3)\ell_s^4}\int r dr
\int_{T^{1,1}}\omega_2\wedge\omega_3
.} According to \normal,\ the integral over $T^{1,1}$ is
$\int_{T^{1,1}} \omega_2\wedge\omega_3=32\pi^3$. We estimate the
integral over the radial direction of the cone with $\int r dr =x
R^2/2.$ Here, $R=V_X^{1/6}$ is the size of $X$ and $x$ is a number
of order one that depends on the details of $X$. Thus, the axion
supported near the vanishing $S^2$ has \eqn\san{F_b=\sqrt{x\over 6}
{R\over \ell_s^2}.} From \aplanck\ and \san,\ we express $R$ and
$M_s=\ell_s^{-1}$ as \eqn\exr{\eqalign{R&=\ell_s
g_s^{1/2}\left(M_P\over F_b\right)^{1/2}\left(x\over
24\pi\right)^{1/4},\cr M_s&={F_b\over g_s^{1/2}}\left(F_b\over
M_P\right)^{1/2}\left(2^5 3^3\pi\over x^3\right)^{1/4}.}}  For $F_b$
in the range preferred by phenomenological considerations $10^9{\rm
GeV}\lesssim F_a\lesssim 10^{12}{\rm GeV},$ with $x=1,$ we have
\eqn\rea{\eqalign{{1.5\times10^5}{\rm GeV}\lesssim &M_s g_s^{1/2}
\lesssim {5\times 10^9}{\rm GeV},\cr 1.6\times 10^4 \ell_s\gtrsim
&{R\over g_s^{1/2}}\gtrsim 5\times10^2 \ell_s.}}

To assess whether the explicit breaking of the Peccei-Quinn symmetry
is sufficiently small, we estimate the actions of instantons that
break it. Generic instantons are Euclidean D1-branes wrapping cycles
of size $\sim R.$ They have very large action $I\sim 2\pi R^2/g_s
\ell_s^2$, since $R>> \ell_s.$ It follows that these instantons
break Peccei-Quinn symmetry negligibly. The main violation of the
PQ-symmetry comes from $D1$-brane instantons that wrap the vanishing
$S^2.$ They have action $I=2\pi
V_Q/g_s\ell_s^2=2\pi/\alpha_{GUT}\sim 157$. Hence, with some help
from low-energy supersymmetry, the PQ-symmetry might be able to
explain the strong CP-problem.

\medskip\noindent{\it Type IIA}

In IIA string theory, we get gauge symmetry by wrapping $D6$-branes
around the small $S^3$ of the deformed conifold.  If the $S^3$ has
radius $r_0,$ the gauge coupling is \eqn\gsi{\alpha_{GUT}={g_s
\ell_s^3\over 2\pi^2 r_0^3}.}  The axions is a four-dimensional
scalar $b$ coming from a zero mode of the RR three-form field $C_3$:
\eqn\ansa{C_3={\omega_3\over 8\pi^2} {b\over 2\pi}.}  $\omega_3$ is
a harmonic three-form on $X$ with a nonzero flux through the
vanishing $S^3$. We approximate it by a harmonic form on the cone,
which is a pullback of the harmonic form $\omega_3$  \clof\ on
$T^{1,1}$. With the help of \normal,\ we normalized the $C$-field so
that the axion $b$ has period $2\pi.$ We find $F_b$ from the general
formula for the decay constant of an RR-axion \gemex\
\eqn\abe{F_b^2={1\over 2\pi\ell_s^2}\left(1\over
8\pi^2\right)^2\int_X \omega_3\wedge\star\omega_3 = {3x\over
4\pi^2\ell^2_s}\ln \left( R\over r_0\right),} where $x$ is a
dimensionless number of order one. In the evaluation of the
integral, we approximated the space $X$ with just the conical
region. The integral over the radial direction of the cone diverges
both for large and small radius. We cut off the large distance
divergence of the integral at the radius $R$ of $X$ and the short
distance divergence at the radius $r_0$ of the vanishing $S^3.$
Since the harmonic three-form is not supported near the vanishing
cycle, the main contribution to the integral comes from the
logarithm $\ln(R/r_0).$ We are justified to neglect the corrections
from the region near the tip of the deformed conifold and from the
region that compactifies the conifold, as long as $R>>r_0.$ If we
assume that the gauge coupling at the string scale is
$\alpha_{GUT}\sim1/25,$ it follows from \gsi\ that $r_0\sim \ell_s.$
But we already know from our estimate \etex\ that $R>>\ell_s$,
whence it follows that $R>>r_0$ and our approximations are
self-consistent.

To find the range of the string compactification parameters that
lead to phenomenologically acceptable axion, we express $M_s$ and
$R$ from \abe\ and \aplanck,\  as \eqn\fino{\eqalign{M_s&={2\pi
F_b\over \sqrt{3x\ln\left(R/r_0\right)}},\cr R&=\ell_s
\left(M_P\over F_a\right)^{1/3}g_s^{1/3}\left(3x
\ln\left(R/r_0\right)\over 2^4\pi^3\right)^{1/6}.}} For $10^9{\rm
GeV}\lesssim F_b\lesssim 10^{12}{\rm GeV}$ and $x=1$, we have
\eqn\scalo{\eqalign{1.4\times 10^9 \GeV\lesssim &M_s \lesssim
1.8\times 10^{12}\GeV,\cr 800 \ell_s\gtrsim &{R\over
g_s^{1/3}}\gtrsim 73\ell_s.}}

\newsec{Type I String Theory}

Just like heterotic string theory, compactifications of type I
string has a model dependent-axion and a number of model-dependent
axions. These come from zero modes of the RR two-form field $C$. The
computations of axion properties in type I string are analogous to
the ones in heterotic string. Here, we will illustrate them in the
case of model-independent axion. For some aspects of axion physics
in Type I orbifold compactifications, see \AngelantonjHI.\

The type I supergravity action is (12.1.34) of \polchinski\
\eqn\typeonea{S={2\pi\over \ell_s^8 g_s^2 }\int d^{10}x
(-g)^{1/2}R-{4\pi\over \ell_s^4}\int \half F_3\wedge\star
F_3-{\sqrt{2}\over 4(2\pi)g_s\ell_s^6} \int \tr F_2\wedge\star F_2.}
where $F_2$ is the $SO(32)$ gauge field strength and $\tr$ is the
trace in the ${\bf 32}$ vector representation of $SO(32).$ We fixed
the normalization of the string coupling $g_s$ using the convention
\eqn\onegravitycoupling{\kappa^2=g_s^2 \kappa_{10}^2=g_s^2
\ell_s^8/4\pi,} and we substituted the gauge coupling from eq.
(13.3.31) of \polchinski:\
\eqn\onegauge{g^2_{Y\negthinspace
M}=2(2\pi)^{3/2}\ell_s^2\kappa=\sqrt{2}(2\pi)g_s\ls^6.} We
normalized the RR-two form, so that its field-strength, (12.1.35) of
\polchinski\  \eqn\filan{F_3=dC-{1\over 16\pi^2} \omega_3(A)} has
integer periods. Thus, our C-field is related to $C_{conv}$ of
\polchinski\ via $C_{conv}=\sqrt{2}\ell_s^2 C.$  From \typeonea\ we
read off the four-dimensional gauge and gravitational couplings
\eqn\onecop{\alpha_{GUT}={g_s \ell_s^6\over \sqrt{2} V_X}\qquad
M_P^2={4\pi V_X\over g_s^2 \ell_s^8}.} Here, we assume usual
embedding $SU(5)\subset SO(10)\subset SO(32)$ of $SU(5)$ in
$SO(32).$

The model-independent axion, comes from a mode of the $C$ field
constant on $X$ with all indices along the Minkowski space. As
explained in section 3 in our study of model-independent axion in
heterotic string theory, the axion decay constant is the inverse of
the coefficient of the $C$-field kinetic energy
\eqn\fain{F_a={\ell_s^2\over \sqrt{4\pi V_X}}={M_P\over
\sqrt{2}}{\alpha_{GUT}\over 2\pi}.} This gives $F_a=1.1\times
10^{16}\GeV$, which is same as the axion decay constant of the
heterotic string model-independent axion. Indeed, under the
heterotic-type I duality, the RR $C$-field model independent axion
of type I string goes into the NS-NS $B$-field model independent
axion of heterotic string theory.

\newsec{ Axion Coupling to Photons}

This concluding section is devoted to a topic in axion physics
that has nothing to do with string theory. \nref\Kaplan{D. B.
Kaplan, ``Opening The Axion Window,''Nucl. Phys. {\bf B260}, 215
(1985).} \nref\Srednicky{M.~Srednicki, ``Axion Couplings To
Matter. 1. CP Conserving Parts, '' Nucl. Phys. {\bf B260}, 689
(1985).} \nref\GKL{H. Georgi, D. B. Kaplan, and L. Randall,
``Manifesting The Invisible Axion At Low Energies,'' Phys. Lett.
{\bf B169}, 73 (1986).} We will reconsider a matter discussed in
section 2, namely the axion-photon coupling. This coupling has a
contribution from short distance  axion physics and another
contribution from low energy QCD strong coupling effects. In
section 2, we evaluated the first type of contribution in a
unified $SU(5)$ gauge theory. To get the physical axion-photon
coupling, one must also evaluate QCD strong coupling effects that
mix the axion with the $\pi^0$. These effects have been previously
determined in \refs{\Kaplan-\GKL} using current algebra
techniques. Here we will provide a new derivation of this coupling
based on the relevant low energy effective Lagrangian.

In section 2, we described the kinetic term and the mass term of
the low energy effective Lagrangian describing pion physics
\eqn\lag{S_0=-{F_\pi^2\over16}\int {\bf{tr}} \l( U^{-1} \p_\m
U\r)\l( U^{-1}\p^\m U \r) + {v\over2}{\bf{tr}}\l(MU+\overline{M}
U^{\dag}\r).} Here $U$ takes values in the group manifold $SU(3)$
and $F_\pi=184\MeV$ is the pion decay constant. $U$ is
conventionally parametrized as \eqn\upi{U(x)=\exp\l({2i\over
F_\pi}\sum_{a=1}^8 \la^a \pi^a(x) \r),} where  $\la^a$ are the
generators of the $SU(3)$ algebra normalized as $\Tr
\la^a\la^b=2\d^{ab}.$ If we integrate out the $s$ quark, which is
much heavier than the $u,d$ quarks, then $U$ is an element of
$SU(2)$ and we take $\la^a=\s^a,\, a=1,2,3$  to be the Pauli
matrices.

\nref\wesszumino{J.~Wess and B. ~Zumino,``Consequences Of Anomalous
Ward identities,'' Phys. Lett. {\bf B37}, 95 (1971).}

The anomalous couplings of pions are summarized in the
Wess-Zumino-Witten term \refs{\wesszumino,\WittenTW}. To write it
down, we let $D$ be a five-dimensional space bounding our
four-dimensional spacetime. We extend $U$ to a map from $D$ to
$SU(3)$. The WZW term is \WittenTW\  \eqn\gterm{\G=-{i
N_c\over2\pi^2\times 5!}\int_D d\Sigma^{ijklm}{\bf} (U^{-1}\p_i
U)(U^{-1}\p_j U)(U^{-1}\p_k U)( U^{-1}\p_l U)(U^{-1}\p_m U),} where
$d\Sigma^{ijklm}$ is the five-dimensional volume element on $D$. The
integrand is $N_c$ times the generator of $H^5(SU(3),2\pi\QZ),$
where $N_c=3$ is the number of colors. The factor of 2 in the
denominator comes from the Bott periodicity theorem \nref\bott{R.
Bott and R. Seeley,``Some Remarks On The Paper Of Callias,''  Comm.
Math. Phys. {\bf 62} (1978) 235.} \bott.\

\subsec{Gauging Electromagnetism}

The WZW action \gterm\ captures the anomalous couplings between
the pions and axions. The couplings of the pion and axion to the
photon can be described by a gauged version of the WZW term. Let
us discuss briefly how one gauges the action $S=S_0+\Gamma$. For
further details, see  \WittenTW.\ The action $S$ is invariant
under the global $U(1)_{\rm EM}$ symmetry \eqn\mira{\d U=
i\e[\CQ,U],} where $\CQ$ is the electric charge matrix of the $u$,
$d$, and $s$ quarks \eqn\qira{
\CQ=\pmatrix{{2\over3}&0&0\cr0&-{1\over3}&0\cr 0 & 0 & -{1\over
3}\cr}.}   We now promote $U(1)_{EM}$ to a local gauge symmetry.
This is accomplished as follows. The kinetic term $S_0$ becomes
gauge-invariant if we replace the ordinary derivatives with
covariant ones \eqn\covar{D_\m=\p_\m+ ie A_\m[\CQ,\ ].} The
$U(1)_{EM}$ gauge field is canonically normalized so it transforms
as $A_\m\rightarrow A_\m-\p_\m \lambda$ under a gauge
transformation $\lambda.$

The WZW term cannot be written in four dimensions as an integral
of a manifestly $SU(3)\times SU(3)$ invariant expression, so the
standard gauging procedure is not applicable to it. Instead, the
gauge invariant generalization of $\Gamma$ can be obtained using
the Noether procedure. A short calculation shows that the desired
modification of $\G$ that makes it gauge invariant is \WittenTW\
 \eqn\inva{\eqalign{\tilde\G=\G-&e\int d^4x A_\m J^\m+
 {ie^2\over24\pi^2}\int d^4x\,
 \e^{\m\n\a\b}\l(\p_\m A_\n\r)A_\a\cr &\times{\bf tr}\big[
 \CQ^2\l(\p_\b U\r)U^{-1}+\l(\p_\b U\r)\CQ^2U^{-1}+\CQ\l(\p_\b U\r)\CQ
 U^{-1}\big],}} where \eqn\jmu{J^\mu={1\over
 48\pi^2}\epsilon^{\alpha\beta\gamma\delta}{\bf tr}\l[\CQ \l(\p_\nu U
 U^{-1}\r)\l(\p_\a U U^{-1}\r)\l(\p_\b U U^{-1}\r)+\CQ \l(U^{-1}\p_\nu U
 \r)\l(U^{-1}\p_\a U \r)\l(U^{-1}\p_\b U \r)\r].}

The reason that we started with three flavors was that this gave a
convenient way to determine the anomalous interactions, which are
summarized in \inva.  Now that we have determined those
interactions, we can for our purposes here omit the strange quarks
and reduce to the case of two flavors.

\subsec{Axion-Photon Coupling}

In eq. \corko, we determined the part of the axion-photon coupling
coming from explicit coupling of the axion to the $SU(5)$ gauge
fields: \eqn\rira{{4r\over3}{\a\over8\pi F_a} a\,
\e^{\m\n\a\b}F_{\m\n}F_{\a\b}={4r\over3\pi}{\a\over F_a} a\,
\vec{E}\cdot \vec{B}.} Here $a$ is the canonically normalized axion
which has periods $2\pi F_a$. Now, we will compute the additional
contribution to the axion-photon coupling due to the mixing of the
axion with the neutral pion. The axion enters the effective action
as a phase of the determinant of the quark mass matrix.  In the
two-flavor approximation, we can take this mass matrix to be
\eqn\pima{M=\pmatrix{ \exp({-ic_u {a/ F_a}}) m_u & 0 \cr 0&
\exp({-ic_d {a/ F_a}}) m_d},} where $c_u+c_d=1$ so that $\det
(M)\propto \exp(-ia).$  The $(\pi^0, a)$ mass matrix comes from
expanding the mass term  \lag\ to quadratic order
\eqn\pimas{\eqalign{\CL_m &={v\over2}{\bf tr}\l(M
U+\overline{M}U^{\dag}\r) \cr &=-{2v\over
F_\pi^2}(m_u+m_d)\l[\pi^0-{a F_\pi\over 2F_a}{c_u m_u-c_d m_d\over
m_u+m_d}\r]^2-{v\over2F_a^2}{m_u m_d\over m_u+m_d} a^2.}} From
\pimas\ we read off the masses of the pion and the axion
\eqn\mapi{m_\pi^2={4v\over F_\pi^2}(m_u+m_d),\qquad m_a^2={m_\pi^2
F_\pi^2\over4 F_a^2}{m_u m_d\over (m_u+m_d)^2}.} The anomalous QCD
contribution to the coupling of axion and pion to two photons comes
from the last term in the gauged WZW term \inva,\ since this is the
only piece quadratic in the $U(1)_{EM}$ gauge field. To find these
couplings,  we perform the $SU(2)_L$ axial rotation
\eqn\maxi{U\rightarrow \pmatrix{\exp(+i c_u{a\over F_a}) &0 \cr 0
&\exp(+ic_d {a\over F_a})}U} and expand \inva\ to first order in
$\pi^0$ and $a.$  The couplings to two photons are \eqn\pina{
-{\a\over 8\pi} \l( {a\over 3 F_a}(4c_u+c_d)+2{\pi^0\over
F_\pi}\r)\e^{\m\n\a\b}F_{\m\n} F_{\a\b}.}

To determine the axion-photon coupling from \pina\ we take a
shortcut and set $c_u m_u-c_d m_d=0$ so that the axion does not mix
with the neutral pion in the axion-pion mass matrix \pimas.\ This
condition together with the constraint $c_u+c_d=1$ determines $c_u=
m_d/(m_u+m_d), c_d=m_u/(m_u+m_d)$. Substituting this into the
axion-photon vertex \pina\ gives the axion-photon coupling
\eqn\pifa{-{r\over 3} {\a\over 8\pi F_a} {m_u+ 4m_d\over m_u+m_d}
a\, \e^{\m\n\a\b} F_{\m\n} F_{\a\b}.} To get the complete coupling
of the axion to two photons we add to this the $SU(5)$ contribution
\rira\ \eqn\tofa{ {r\over 3}{\a\over 8\pi F_a} \l(4- {m_u+4m_d\over
m_u+m_d}\r)a\, \e^{\m\n\a\b} F_{\m\n} F_{\a\b}=-{r\a\over \pi F_a}
{m_u\over m_u+m_d}a\, \vec{E}\cdot \vec{B}.} The coupling \tofa\
depends on the mass ratio of of the light quarks and it vanishes for
$m_u=0.$ In the nature $m_d/m_u\simeq 1.8/1,$  hence the axion to
two photons coupling is suppressed by a factor of $\sim 4$ compared
to the $SU(5)$ result \rira\ alone without the QCD correction.

\bigskip

P.S. would like to thank S. Kachru, J. McGreevy and M. Dine for
useful discussions. P.S. and E.W. would like to similarly thank P.
Steinhardt.  We also thank C. Kokorelis and M. Porrati for helpful
comments on an earlier draft.  This work was supported in part by
NSF Grant PHY-0070928 and PHY-0244728, and the DOE under contract
DE-AC02-76SF00515.

\listrefs
\end